# *Spin–lattice coupling enables adaptive adsorption in magnetically-driven electrocatalysts*

Arnold Gaje[1†], Lulu Li[2†], Felipe A. Garcés-Pineda[2], Camilo A. Mesa[3*], Ghazaleh Abdolhosseini[2,4], Aditya K. Kushwaha[1], Dora Zalka[5], Elzbieta Trzop[1,7], Nicolas Godin[1], Raffaella Torchio[5], María Escudero-Escribano[3,6], Eric Collet[1,7], Sixto Giménez[8], Niels Keller[1], José Ramón Galán-Mascarós[2,6], Núria López[2*], Ernest Pastor[1,7*]

[1] *CNRS, Univ. Rennes, Institut de Physique de Rennes, Rennes, France.*
[2]*Institute of Chemical Research of Catalonia (ICIQ-CERCA), The Barcelona Institute of Science and Technology (BIST), Av. Països Catalans, 16, 43007 Tarragona, Spain*
[3]*Catalan Institute of Nanoscience and Nanotechnology (ICN2), CSIC, Barcelona Institute of Science and Technology, UAB Campus, 08193, Bellaterra, Barcelona, Spain*
[4]*Departament de Química Física i Inorgànica, Universitat Rovira i Virgili, Marcel. lí Domingo 1, Tarragona 43007, Spain*
[5]*European Synchrotron Radiation Facility, Grenoble 38000, France*
[6]*ICREA, Passeig Lluís Companys 23, 08010 Barcelona, Spain*
[7]*CNRS, Univ Rennes, DYNACOM (Dynamical Control of Materials Laboratory) - IRL2015, The University of Tokyo, Tokyo, Japan.*
[8]*Institute of Advanced Materials (INAM) Universitat Jaume I, 12006, Castelló, Spain*

* camilo.mesa@icn2.cat, nlopez@iciq.es, ernest.pastor@cnrs.fr

† Contributed equally

**A major challenge in electrochemistry is to decouple the reactive intermediates of a catalytic cycle to optimise their energies independently. During the oxygen evolution reaction (OER), such energy interdependence results from the need to generate multiple adsorbates at the same site and sets the minimum overpotential. Here, we show that an external stimulus, such as a magnetic field, can relax the scaling relationships between intermediates during the OER. Spectroscopic measurements and Density Functional Theory simulations in Ni-Fe oxyhydroxides reveal that applying a magnetic field alters surface chemisorption and injects structural flexibility at the interface. We interpret these observations as a consequence of stimulated changes in the spin-lattice coupling, which allow access to quasi-degenerate oxygenated intermediates that modulate the reaction energy demands. Our findings redefine the scaling limitations as state-projected rather than intrinsic and establish external stimulation as a strategy to navigate multi-state energy landscapes in electrocatalysis and sensing applications.**

Photo-electrocatalytic cycles are described as a series of proton and electron transfer steps that generate reactive intermediates at the catalyst surface[1]. A big challenge in catalyst optimisation is that the adsorption energies of such intermediates are linearly dependent, and cannot be optimised individually to maximise the efficiency of each step in the cycle[2,3].

Such linear-scaling relationship (LSR) limitations are well known for metal surfaces[4]. However, increasing evidence suggests that transition metal oxides (TMOs) may offer pathways to overcome this constraint[5–7]. The difference is that while in metals the density of states typically remains unaltered upon electronic polarisation, TMOs can exhibit pseudo-capacitive behaviour, enabling charge storage and sustaining changes in their bonding structure when electrically biased,[8,9] as observed in in-situ optical and X-ray spectroscopy studies[8,10–13]. This process is akin to the accumulation of redox equivalents in the natural photosynthetic system and implies electronic and atomic structure changes under polarisation[14,15]. As a consequence, in TMOs, the redox intermediates can be considered as polaronic states, in which electronic and atomic structures are strongly coupled[16].

This phenomenon has two key consequences. First, each type of reactive intermediate may be associated with a metal in a different oxidation state. For example, in iridium oxide, the different oxygenated intermediates formed during the oxygen evolution reaction (OER) have been associated with changes in the oxidation state of Ir, from Ir(III) to Ir (V)[9,17]. Similar observations have been made in other TMOs, such as Co and Ni oxides[15,18–20].

Second, the charge, orbital, lattice, and spin degrees of freedom are strongly coupled, and change along the reaction pathway as the oxidation state and the coordinating or bonding structure of the intermediates evolve[21]. For example, the addition/removal of a d-electron during a reaction step may render a metal centre magnetically active or lift orbital degeneracies through Jahn-Teller distortions[22–24].

The coupling between the different degrees of freedom of a solid is exploited in condensed matter devices to induce technologically relevant functions, like magnetic or superconducting states[25–27]. In electrochemistry, exploiting this coupling could offer a path to energetically decouple the reaction intermediates and break LSRs by using external stimuli. For example, during the OER, different oxygenated intermediates are generated upon electrical polarisation (**Fig. 1a**). The application of an additional stimulus may further alter the electronic and bonding structures along the reaction pathway, modulating the energy of adsorption of each intermediate separately or even favouring the formation of new intermediates that would be otherwise inaccessible under regular potentiostatic control (**Fig. 1b-c**).

However, achieving such selective control of chemisorption in oxides remains challenging. Experimentally, it is difficult to characterise surface-bound intermediates over a broad range of applied potentials on structurally complex electrodes. Meanwhile, conventional theoretical approaches based on Density Functional Theory (DFT) do not often capture the diversity of accessible electronic and structural states, limiting the ability to model adsorption behaviour accurately.

Herein, we evaluate the feasibility of controlling the formation of reactive intermediates on demand. We observe that magnetic stimulation of a Ni-Fe oxide modulates the energetics of the intermediates during the electrocatalytic OER. We interpret this data as a magnetically-driven change in the chemisorption and a relaxation of linear-scaling relationships. We selected the Ni-Fe oxide because it is a benchmark OER electrocatalyst in alkaline media, its surface speciation can be studied spectroscopically[18,28–32], and its OER faradaic current increases under magnetic field application[33,34].

We perform DFT simulations that consider thermodynamic stability across different spin and oxidation states and find that a range of intermediates around the Ni, Fe and O atoms is accessible. Many of these states are energetically similar, yet they have distinct atomic and electronic structures which respond differently to geometrical perturbations induced by external stimuli. DFT also predicts that subtle electronic differences in the adsorbates enable new reaction pathways that bypass energy barriers. Spectroscopic measurements validate that the external field triggers changes in the adsorbate populations. Our data show that magnetic stimulation can decouple intermediates and affect lateral interaction between oxygenated adsorbates, leading to current enhancements. The framework we discuss may apply to other stimulation methods that interact with the relevant microscopic degrees of freedom.

## Results

### Electronic and geometric diversity

Ni-Fe oxides are well known to evolve at the surface into a layered oxyhydroxide as the OER active phase[35–37]. Thus, we first use DFT simulations to evaluate the thermodynamic stability and structural features of $Fe_xNi_{(1-x)}OOH$ across different spin and oxidation states distributions to understand the nature of the surface (**Fig. 2a-d;** see Supplementary Note 1). To this end, the surface is described using two complementary NiFeOOH models: a substitutional Fe-doped NiOOH and an Fe-adatom NiOOH surface (**Fig. S1**), capturing distinct Fe coordination motifs. NiOOH is also calculated and shown in **Fig. S2**. This approach allows us to systematically disentangle the contributions of different structural and electronic factors, including Fe incorporation method (substitution versus adatom), magnetic state configuration, and oxidation state distribution.

To illustrate the distinct oxidation types that can form at the active site, we classify representative configurations in **Fig. 2a**. In the first deprotonation step (DP1), oxidation can localize on three possible redox-active centres: Ni, Fe, or O, denoted as Ni-ox, Fe-ox, and O-ox, respectively. Upon further deprotonation (DP2), more complex oxidation patterns can emerge, including metal-centred oxidation with inter-site charge redistribution (Fe-Ni and Ni-Ni), metal-oxygen coupled oxidation (Fe-O and Ni-O), and oxygen-centred oxidation states (O-O), reflecting increased charge delocalisation.

Our analysis reveals a broad distribution of accessible oxidation states with relative stabilities varying under both high-spin (HS) and low-spin (LS) initial states of the active site (**Fig. 2b** and **2c**). These variations depend sensitively on the reaction stage and underlying magnetic configuration. Such electronic-state multiplicity is fundamentally distinct from the behaviour of metallic surfaces commonly used in electrochemical models and is a defining feature of TMO electrodes [38–42]. Notably, despite nearly identical $\Delta G_{*OH}$ references within each model, the $\Delta G_{*O}$ spans a wide energetic manifold (**Fig. S3** and **S4**), resulting in a non-single-value mapping between these two intermediate states. Such behaviour directly violates the single-valued functional correspondence underlying conventional adsorption scaling relations.

A general stability trend emerges: oxidation solely at metal sites (Fe-Ni or Ni-Ni) is the most stable, followed by coupled sites (Ni-O ≈ Fe-O), and finally oxygen-centred (O-O) species. The higher formation energy of O-O type oxyl radical species suggest that these require additional stabilisation, likely rendering their formation the potential-determining step during the OER under certain conditions. We find that even within the same oxidation state, a dispersed energy range exists, which arises from different possibilities of magnetic distribution among the metal sites.

Such configurational diversity is further exemplified in the bonding structure. We observe variations in bond lengths across oxidation states (**Fig. 2d**), pointing to distinct local geometries depending on the oxidation site. For 1O* intermediates during the OER, Ni-centred (Ni-ox) and Fe-centred (Fe-ox) species show relatively similar M-O bond lengths in the range of ~2.08-2.12 Å. In contrast, O-centred (O-ox) species, corresponding to $O^{\bullet}$, display a bond extending up to 2.30 Å, reflecting significant elongation with the O-M bond when oxygen accommodates the oxidation. Additional bond length information is provided in **Fig. S5**.

Comparison of different spin states at the Fe site (**Fig. 2d**) further reveals that for the same 2O* Fe-Ni oxidation type, the M-O bonds remain similar between intermediate-spin (IS) and HS states — approximately 2.00-2.07 Å. One direct consequence of the minor geometric differences between spin arrangements is that spin-state interconversion may occur with minimal structural redistribution. Theoretically, during the reaction, such quasi-degenerate configurational flexibility could result in multiple spin configurations becoming readily accessible along the reaction pathway.

To evaluate how this diversity impacts the reaction energetics, we computed free energy profiles for HS and LS states separately and their combination (**Fig. 2e).** The OER mechanisms are estimated at pH = 13 and U = 1.23 V. The most favourable HS reaction pathway identified under these conditions yields a theoretical overpotential of 0.44 eV, with oxidation at the Ni site in the first step being particularly favourable (**Fig. 2e orange**).

Under LS conditions (**Fig. 2e blue**), the energy landscape shifts markedly, offering broader thermodynamic accessibility and lower overall barriers. The LS state generally stabilises intermediates more effectively, resulting in a lower minimum theoretical overpotential of 0.26 eV. However, this advantage is counterbalanced by the substantial energy penalty associated with transitioning Fe sites from HS to LS states, which can reach up to 0.82 eV.

While the individual spin configurations (HS or LS) impose a high energy cost (**Fig. S3**), their combination identifies an optimal pathway with a theoretical overpotential as low as 0.29 eV (**Fig. 2e red**). Such reduced overpotential is a strong predictor that an external stimulus affecting the spin state could energetically decouple the reaction intermediates, thus circumvent linear-scaling relationships and offering a more energetically favourable transition along the OER coordinate.

We next test whether it is possible to act on the spin state of the electrode at different applied potentials with the use of a magnetic stimulus. We evaluate if such a perturbation can trigger the emergence of new and energetically different adsorbates and inject flexibility into the reaction coordinate.

**Magnetic control of the electrochemical response**

We start by measuring the electrochemical response of the Ni-Fe electrocatalyst with and without a magnetic field to assess the nature of the magnetic enhancement (see Supplementary Note 5). The current-voltage response in the absence of the magnetic field (**Fig. 3a**) is characterised by a redox wave around 1.5 $V_{RHE}$, associated with the oxidation of the Ni centres, followed by rapid current growth at more oxidative potentials due to the OER. This response is typical of Fe-doped Ni oxyhydroxides[18,43]. Upon application of a constant magnetic field of $H$ = 0.7T, we observe an increase in the current density at applied potentials greater than 1.75 $V_{RHE}$, suggesting a magnetic enhancement of the OER, in agreement with previous studies of related systems.[33,34]

We find that the change in the current density depends on the applied magnetic field strength and polarity (**Fig. 3b** and **Fig. S10**). We measure a rapid current change up to 0.4 T, followed by a tendency to saturation at higher fields and a nearly symmetric response at positive and negative polarities. Moreover, when scanning from positive to negative fields (and reverse), we observe that the current minimum occurs at $H$ = + 0.1 T (– 0.1 T in the reverse scan) instead of $H$ = 0 T. Such a difference of the current minima between forward and reverse scans is reminiscent of the coercive field characteristic of ferro- or ferrimagnetic materials, which reflects the magnetic hysteresis loop and the remanent magnetisation in zero applied field, similar to "memory-type effects".[44,45]

To test the link between the current and the electrode's magnetisation, we compare the electrochemical current density with the magneto-optical Faraday (MO) response of the sample, which tracks the magnetic order of the oxide layer. **Fig. 3b and 3c** show a strong correlation between these two observables. Specifically, the applied field at which the current density tends to saturation matches the magnetic field of complete magnetisation. This occurs both at positive and negative polarities. Moreover, the MO signal exhibits a ca. 0.1 T coercive field, similar to that observed in the electrochemical response (**Fig. 3c**, see Supplementary Note 3 for a description of the magnetooptical response).

These results suggest that the magnetic properties of the catalyst influence its electrochemical response. The hysteretic behaviour of the current on the applied magnetic field indicates that changes in the magnetic structure of the electrode are important to control its electrochemical response. While magnetic effects are complex to disentangle, these observations suggest that the external stimuli can act on the reaction kinetics, as hypothesised in previous studies[34,46–50].

These observations are also in agreement with our simulations that suggest a more energy-efficient OER pathway upon control of the spin-state. We next evaluate whether such phenomena occur due to a change in the reaction intermediates and a breaking of linear-scaling relationships.

**Redox transitions upon external stimulation**

To test if the magnetic stimulus induces changes in the population of the intermediates, we perform UV-VIS spectroelectrochemistry measurements. In these experiments, the optical changes of the sample are recorded as a function of the applied potential[51,52]. In transition metal-containing systems, such optical response is associated with changes in the d-orbital configuration that occur as intermediates are formed. Specifically for Fe-/Ni-based systems and other TMO electrodes, previous works have shown that spectroelectrochemistry in the UV-Vis can track and quantify the redox species involved in the OER[9,18,20,53–56]. For our measurements, we developed an *in-situ* set-up based on a phase-sensitive detection **(Fig. 4a**, see Supplementary Note 5c), which allows us to compare the electrode response with and without an applied magnetic field.

**Fig. 4b** shows the differential UV-VIS spectrum under different experimental conditions (see also **Fig. S11**). In agreement with previous reports, the signal increases with increasing applied potential and exhibits higher intensity at shorter wavelengths. As discussed below, such a response has been linked to the formation of the oxidative intermediates involved in the OER. We observe that, at oxidative potentials, the application of a magnetic field causes further changes in the optical response compared to the same conditions without a magnetic field, suggesting that the stimulus modulates the electronic state of the catalyst.

To explore a wider voltage range, we recorded the absorbance-voltage characteristics over a large voltage window at ~ 540 nm. In the absence of a magnetic field, the response is characterised by a sharp change in absorbance around 1.5 $V_{RHE}$, followed by a slower signal growth that subsequently plateaus and remains saturated within the measured range (**Fig. 4c**). When the applied potential is reversed, the signal remains saturated at high applied potentials and decreases sharply around 1.4 $V_{RHE}$ (**Fig. 4d**).

Upon application of a constant magnetic field of $H$ = 0.7T, we measure a similar behaviour at low applied potentials (**Fig. 4e**). However, at higher potentials, we observe a further change in optical density around 1.7 $V_{RHE}$. Upon reversing the applied potential (**Fig. 4f**), we measure a constant signal decrease with a sharp absorption around 1.65 $V_{RHE}$ and a second drop around 1.4 $V_{RHE}$. Strikingly, the new absorption features correlate with the magnetic enhancement in current density observed in **Fig. 3a**.

Without a magnetic field, our differential optical response agrees with previous measurements of Fe-Ni-oxyhydroxides for OER. Specifically, large optical changes at low voltages, which correlated with the redox wave, have been assigned to the Ni(II)/Ni(III) oxidation and deprotonation of the surface[18,29]. The absorption increase at more oxidative voltages than 1.45 $V_{RHE}$ was associated with further deprotonation of surface hydroxyl groups and the formation of reactive oxo species that participate in the rate-determining step of the reaction[18,31,53,57]. These assignments also align with previous XAS studies that revealed the presence of highly oxidised Ni at OER potentials[29,58]. Our measurements further indicate that these transformations may be reversed if the applied potential is scanned in a reducing direction.

Under magnetic stimulation, we observe a further increase in absorbance. Within our broad range of applied voltages, we observe a pronounced signal increase at voltages more oxidative than the OER onset that correlates with the current enhancements. These observations point towards a further change in the range of accessible intermediates under magnetic stimulation.

While our UV-Vis measurements lack chemical sensitivity to inform on the nature of the intermediates, the large voltage and optical signal range accessible in our measurements allows quantitative analysis of the response to estimate thermodynamic values that can be contrasted with DFT simulations.

**Optically-determined surface coverages and binding energies**

The sequential change in the optical response with different saturation regimes we measure is akin to those observed in transition-metal containing systems involving a voltage-dependent generation of reaction intermediates[51]. Moreover, the appearance of successive optical transitions at close applied potentials is reminiscent of the response of systems in which multiple adsorbates have diverse binding modes[59].

To estimate thermodynamic parameters that can be compared with the DFT simulations, we use the absorbance difference as a function of potential to construct electro-adsorption isotherms – a relationship between the density of adsorbed species and the electrode potential. This analysis rests on the assumption that the signal change monitored spectroscopically is directly proportional to the surface concentration of oxidised species, and has been recently applied to a range of TMOs[9,20,59,60] (See Supplementary Note 8 for a discussion on this approximation). This approach allows contrasting the spectroscopic data with electro-adsorption isotherms. Here we test the Langmuir model —that assumes independent adsorbates— and the Frumkin model —that accounts for attractive/repulsive interactions between adsorbates.

In our experiments, we measure signals in the order of 40 m∆OD at oxidative potentials (**Fig. 4c-d**) similar to other reports in nickel oxides[18,34,53,54,56]. However, in contrast to previous studies, our measurements extend over a larger voltage range and capture the attainment of signal plateaus at high voltages, corresponding to a saturation of the density of oxidised species. This signal saturation, which previously could only be postulated, is indicative of the maximum concentration of oxidised species and thus can be approximated to the full coverage ($\theta$) of a redox state.

We focus our analysis on the first two redox transitions, as signal plateaus can be identified within our measured voltage range. We ascribe the transitions from one plateau to another as the transition between coverage regimes. In this representation, in the absence of a magnetic field (**Fig. 5a-b**), two changes in surface coverage can be estimated: at ~1.5 $V_{RHE}$ and ~1.6 $V_{RHE,}$ respectively.

The first transition aligns with the redox wave assigned to the oxidation from Ni(II) to Ni(III)[43]. The second transition has been associated with the deprotonation of M-OH centres and formation of oxo species ($O^*$)[18]. Notably, we observe that the growth of this transition is gradual, spanning up to 2.0 $V_{RHE}$ in the forward scan (**Fig. 5a**), while in the reverse scan, the transition is sharper (**Fig. 5b**), indicating a different energy pathway. In contrast, the sharpness of the Ni(II)/Ni(III) transition is similar between the forward and reverse scans, suggesting a similar transformation. We attribute such behaviour to the polaronic nature of the active sites with strong charge-lattice coupling. Specifically, the similar response in the forward and reverse scans of the first redox transitions suggests minimal structural rearrangements and a primarily local redox process. In contrast, the differences in the oxo transformation point towards larger structural changes, introducing a stronger path dependence. This is in agreement with the observation of changes in the metal coordination environment at OER potentials in EXAFS studies of Ni-Fe oxides and suggests that the charge and lattice degrees of freedom play a key role[61,62].

Within the surface coverage approximation, we find that the first redox transition can be well fitted to a Langmuir model. In contrast, during the forward scan, the second transition due to oxo formation is best fitted to a Frumkin isotherm, which takes into account changes in the surface coverage due to lateral interactions between adsorbates[20,59,60]. We obtain an interaction parameter of ~0.3 eV (a comparison of the different fit models is provided in Supplementary Note 8).

This interaction energy is comparable to that reported for other TMOs based on Ir and Co, where adsorbate interactions control the OER[9,20]. Our results suggest that adsorbate-adsorbate interactions also play a significant role in the OER mechanism in Ni-Fe oxides. Notably, we do not observe such a strong influence of oxo-oxo interactions in the reverse scan, suggesting this contribution is more relevant during the charging process and pointing towards an important role of the lattice in controlling the interactions.

In the presence of a magnetic stimulation, we measure a similar growth of the Ni(II)/Ni(III) transition, suggesting this state is not significantly affected by the magnetic field (**Fig. 4c**). However, we also observe some key differences. First, the surface coverage due to oxo formation, around 1.5 $V_{RHE}$, exhibits a significantly sharper response both in the forward and reverse scans. We note that the sharpness of this transition renders its detection more complex, as the number of data points available is reduced; within our resolution, we observe that a Frumkin fit with a very small (attractive) interaction energy represents the data well.

Second, an additional surface coverage emerges around 1.75 $V_{RHE}$, approximately 250 mV at potentials more positive than the Ni(II)/Ni(III) redox wave. This transition is not detected in the absence of a magnetic field and can be well fitted to a Langmuir model.

The difference in the estimated electro-adsorption isotherms with and without a magnetic field suggests that the magnetic stimulation can influence the electrode by altering the lateral interactions between intermediates and by enabling new redox states at high applied potentials. This aligns with our DFT simulations that predict the stabilisation of oxyl intermediates (M-O*) in the fully deprotonated metal centres, and a much broader distribution of intermediate configurations under magnetic stimulation. We find that the predicted energy barrier of 0.25-0.3 eV between the protonated and magnetically stabilised oxyl states (**Fig. 2e**), is in close agreement with the experimentally measured redox transition at 1.75 $V_{RHE}$.

**Discussion and conclusion**

Our results suggest that the different degrees of freedom available in a TMO, such as spin state and lattice, can be leveraged by an external perturbation to trigger new chemisorption and modulate the interactions between adsorbates, thereby changing the surface energy and the reaction pathways.

Our data show the magnetic field renders a broad range of adsorbates accessible, preferably stabilising oxyl intermediates (M-O•) and oxygen-centred oxidations. These species have localised spin density on the oxygen ligand, which reduces electronic delocalisation and minimises orbital overlap between neighbouring active sites, likely suppressing and modulating lateral repulsive interactions between surface-bound intermediates.

Most importantly, we observe that the stabilisation of new intermediates by the external perturbation disrupts the reaction pathway, leading to a current enhancement bypassing the linear scaling relationships that limit traditional OER mechanisms. We propose that instead of proceeding through a fixed sequence of proton-coupled electron transfers, the system can exploit mixed-spin states and form the O-O bond directly through coupling between two O(oxyl) species.

These findings emphasise the versatility of magnetic perturbations to enhance electrocatalytic processes. In addition to effects such as spin selectivity and spin polarisation in charge transfer and transport or dynamical electrolyte effects, the external field can also act as a lever to control the dynamics of intermediates. Scrutinising all plausible microscopic parameters, their tuneability and their complementarity will be fundamental to establishing the basics of magneto-electrocatalysis[50].

Our results have some key implications. Theoretically, they suggest that reactivity models need to account for subtle differences in oxidation, spin and geometrical configurations and thus for the multiple configurations that the system can adopt. The resulting broad distribution of intermediates can inject flexibility into the reaction coordinate, enabling low-energy pathways which otherwise could go unnoticed (**Fig. 2e**). The development of broadband UV-Vis, IR/Raman and X-ray experiments under both electrochemical and external stimulation will be needed to map the full electrochemical richness of the stimulated surface to validate theoretical models.

Experimentally, our results suggest that the degrees of freedom that d-orbital configurations offer can be exploited by selective perturbations to change the properties of the surface on demand. The use of targeted stimuli to tune the properties of TMOs is currently a central goal in condensed matter research, where perturbations are used to induce new phases of matter with exotic functionality[26,27,63,64]. Similar strategies could find applicability in technologies where governing the surface energy via chemisorption is crucial, such as electrochemical selectivity control, sensing devices or thin-film growth methods.

**Competing Interests**

The authors declare no competing interests.

**Data availability**

All data shown in this study are included in this article (and its Supplementary Information files). DFT structures and computational details are available in the ioChem-BD repository at https://iochem-bd.iciq.es/browse/review-collection/100/114478/bf71bfb8a255eef85e1ff9b1

**Author contributions**

E.P. and C.A.M. conceived the Project. E.P. supervised the Project. A.G. conducted the electrochemical and optical experiments. L.L. conducted DFT calculations under the supervision of N.L. N.K. developed the magneto-optics experimental setup and supervised magnetic characterisation measurements. N.G. assisted with experimental developments. F.A.G.P, G.A., and J.R.G.M developed and characterised the magnetically active electrocatalysts. A.G, A.K.K, D.Z., E.T, R.T, E.P. and C.A.M. performed X-ray experiments. M.E.E, E.C., and S.G. contributed to articulating key concepts. A.G, L.L, N.L. and E.P. wrote the paper with input from all authors.

**Acknowledgements**

SG thanks support from the project PID2023-152771OB-I00, funded by MICIU/ AEI/10.13039/501100011033/, by "ERDF A way of making Europe". EP and AG thank the support of Rennes Métropole and the CNRS and the French Agence Nationale de la Recherche (ANR), under grant ANR-22-CPJ2-0053-01. Funded/Co-funded by the European Union (ERC, PhotoDefect, 101076203). Views and opinions expressed are those of the authors only and do not necessarily reflect those of the European Union or the European Research Council. Neither the European Union nor the granting authority can be held responsible for them. NL and LL acknowledge the support from the Spanish Ministry of Science and Innovation (PID2024-157556OB-I00 funded by MICIU/AEI/10.13039/501100011033/ FEDER, UE). ICIQ is funded by the CERCA programme/Generalitat de Catalunya and is also supported by the Severo Ochoa Excellence Accreditation (CEX2024-001469-S funded by MCIU/AEI/10.13039/501100011033). LL also thanks The European Union's Horizon Europe research and innovation program through the Marie Skłodowska-Curie Actions grant (ADAMox, 101149049) and European Innovation Council Project (Superval, 101115456). The Barcelona Supercomputing Center (BSC-RES) provides generous computational resources. C.A.M. acknowledges the Ramón y Cajal program (RYC2023-045597-I) funded by MICIU/AEI/10.13039/501100011033 and FSE+. The ICN2 is funded by the CERCA programme/Generalitat de Catalunya and is also supported by the Severo Ochoa Centres of Excellence programme, Grant CEX2021-001214-S, funded by MCIU/AEI/10.13039.501100011033. G.A. thanks MCIN/AEI/ 10.13039/501100011033/ for a FPI fellowship (PRE2022-101393). We acknowledge the European Synchrotron Radiation Facility (ESRF) for providing the beamtime. This work is based on experiments performed at the ESRF under proposal CH-7692 under the DOI: https:// doi.org/10.15151/ESRF-ES-2297576860.

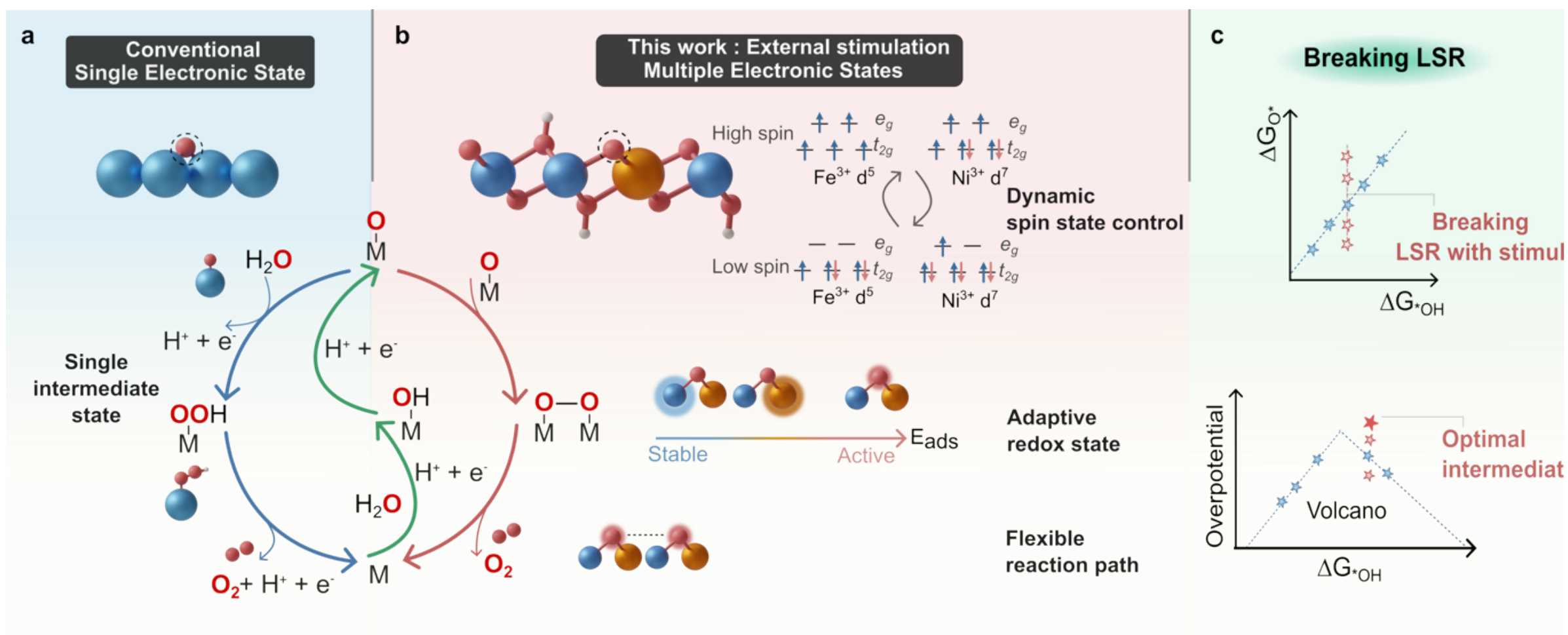

**Figure 1 | Stimulated electrocatalysis in transition metal oxides**. Schematic comparison between conventional electrocatalysis described by a single electronic state and the present concept invoking multiple accessible electronic intermediate states. (**a**) Under an applied anodic potential, surface deprotonation generates oxygenated intermediates (*OH, *O), typically treated within a single-state framework. (**b**) In this work, an additional selective external stimulus (magnetic field) enables access to multiple spin and electronic states of the OER intermediates, dynamically modifying adsorption properties and reactivity. The external stimulation expands the accessible energy landscape of reactivity process, stabilizing otherwise high-energy configurations and activating alternative reaction channels. (**c**) Conceptual energy diagrams highlighting how conventional scaling relations constrain activity to a single volcano optimum, while accessing multiple electronic states allows deviation from linear scaling, enabling optimized binding of key intermediates and potentially breaking linear scaling relations (LSR).

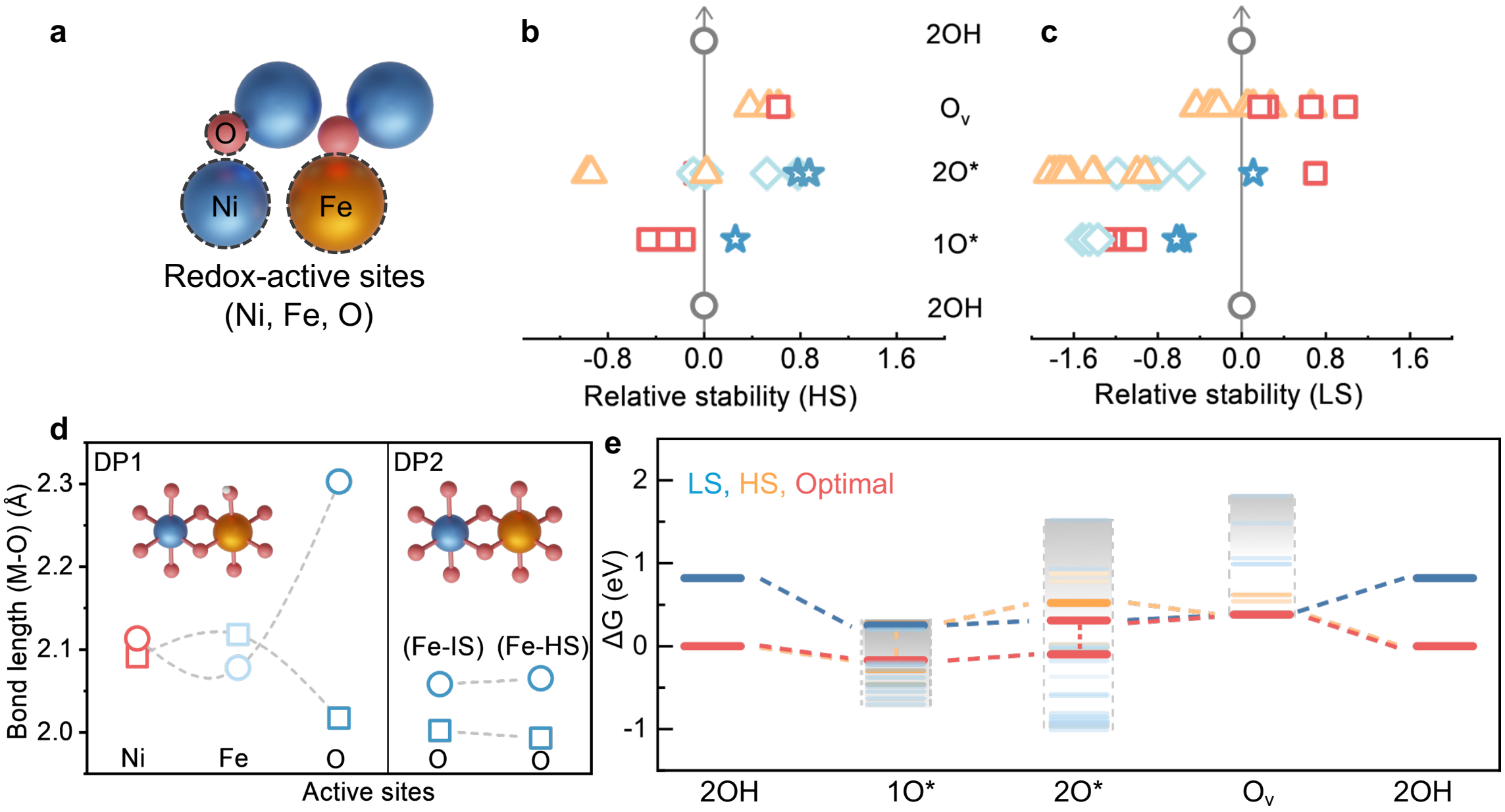

**Figure 2│ Theoretical exploration of the electronic and geometric diversity of FeNiOOH along the OER pathway. (a)** Modelling structure of the NiFeOOH model, highlighting the active Ni, Fe and O sites. (**b, c**) Relative stability of possible oxidation states associated with different reaction intermediates during OER in (b) HS and (c) LS configurations. Colours indicate the localisation of the oxidation centre. For the 1O*, oxidation may localise on Ni (red), Fe (light blue), or O (dark blue). For the 2O*, additional mixed configurations appear, including Fe-Ni coupled oxidation (orange), together with Ni-, Fe-, and O-centred states. **(d)** Variation of Ni-O (squares) and Fe-O (circles) bond lengths as a function of oxidation states and active sites. **(e)** Free energy profile of the OER pathway on NiFeOOH. Blue dashed lines denote pathways within low-spin states; orange dashed lines correspond to high-spin pathways; red dashed lines represent the globally optimal route obtained by allowing spin-state crossover between LS and HS configurations at each intermediate step.

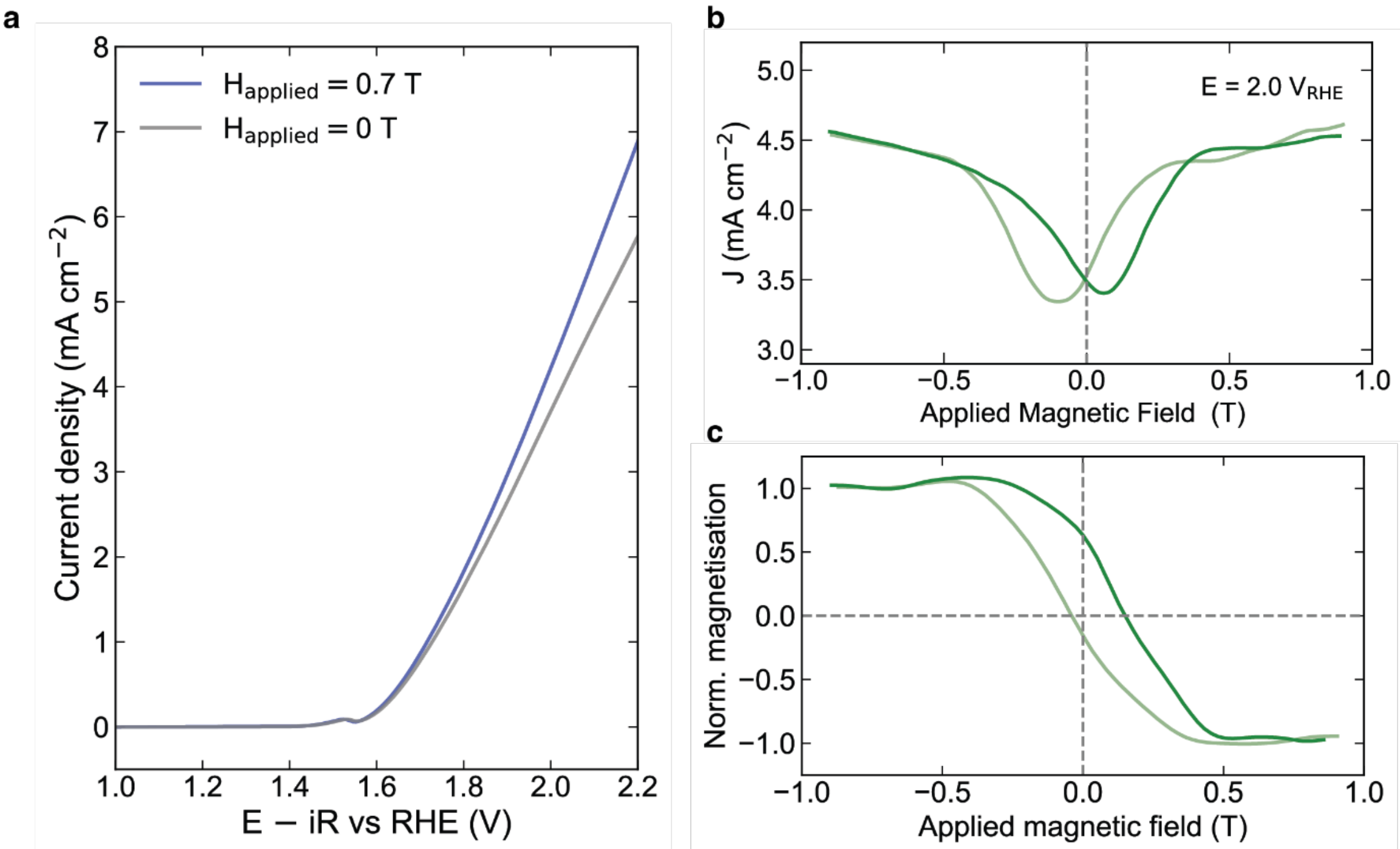

**Figure 3| Enhancement of OER activity under applied magnetic field. (a)** Current-voltage characteristics of the Ni-Fe oxide at a scan rate of 10 mV/s in 0.1 M KOH with and without applied magnetic field. (**b)** Current density enhancement at 2.0 $V_{RHE}$ as a function of applied magnetic field strength. **(c)** Magnetic hysteresis loops of the as prepared catalyst, at 295 K based on magneto-optical Faraday spectroscopy measurements at λ = 542 nm. See Supplementary Information for extended dataset and magnetic signal analysis.

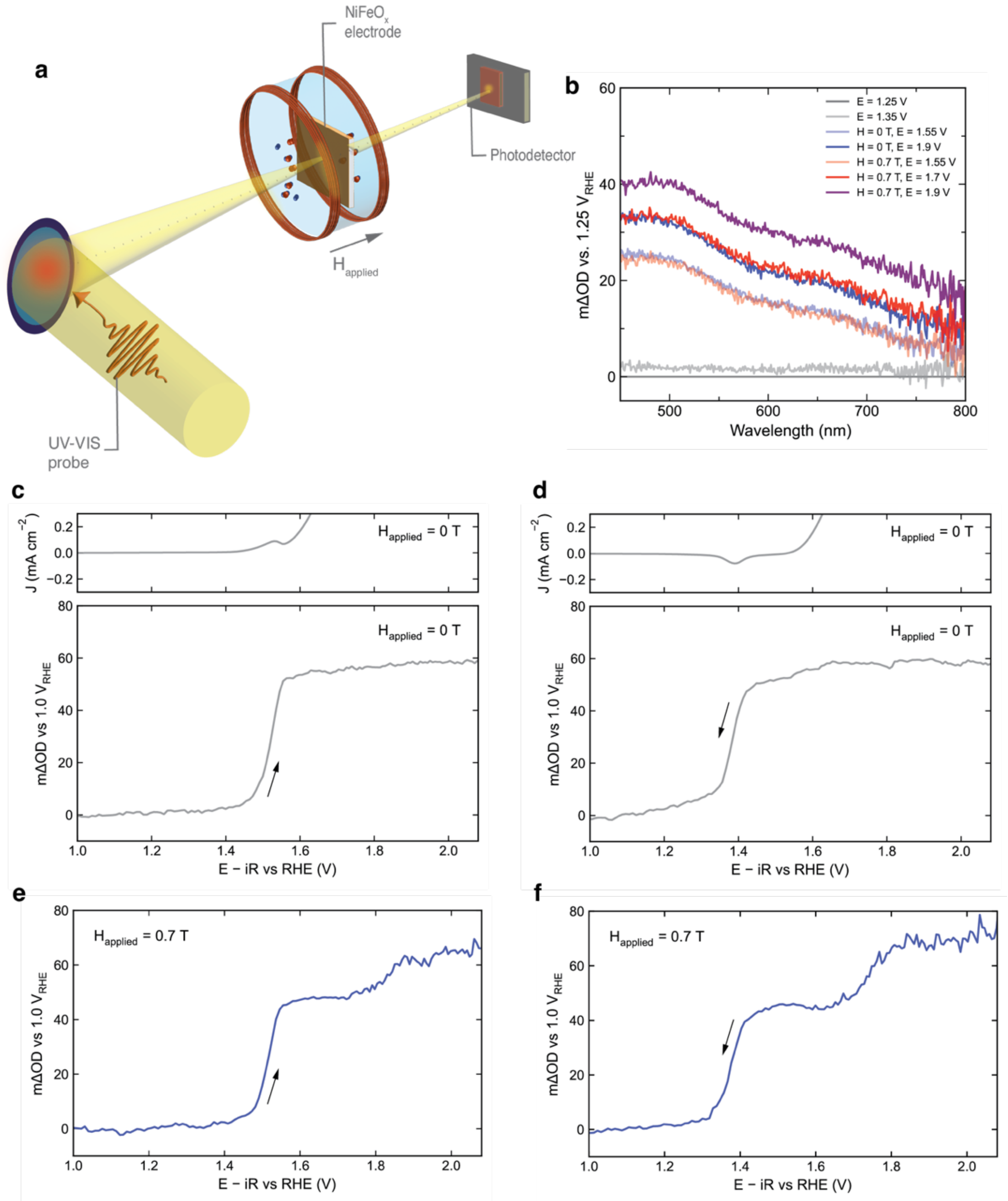

**Figure 4│ Operando UV-Vis Spectroelectrochemistry measurements. (a)** In situ spectroelectrochemistry setup used to monitor effect of magnetic stimulation on the OER. (**b**) Representative differential absorption spectra at different applied potentials with and without magnetic field collected under steady state conditions (see Fig. S11 and Supplementary Note 6 for an extended potential range). **(c-f)** Differential absorption at 542 nm measured at scan rate of 10 mV $s^{-1}$. (c) and (d) show data without magnetic stimulation for the forward and backwards scan respectively alongside the corresponding current density. We observe a sharp change in the optical signal that correlates with the Ni(II)/Ni(III) redox transition. (**e**) and (**f**) show data with a constant magnetic stimulation at 0.7 T for the forward and backwards scan respectively. A further change in the differential signal is observed at high applied potential. The data was collected in 0.1M KOH. See Supplementary Notes 5 and 7 for extended datasets and experimental details.

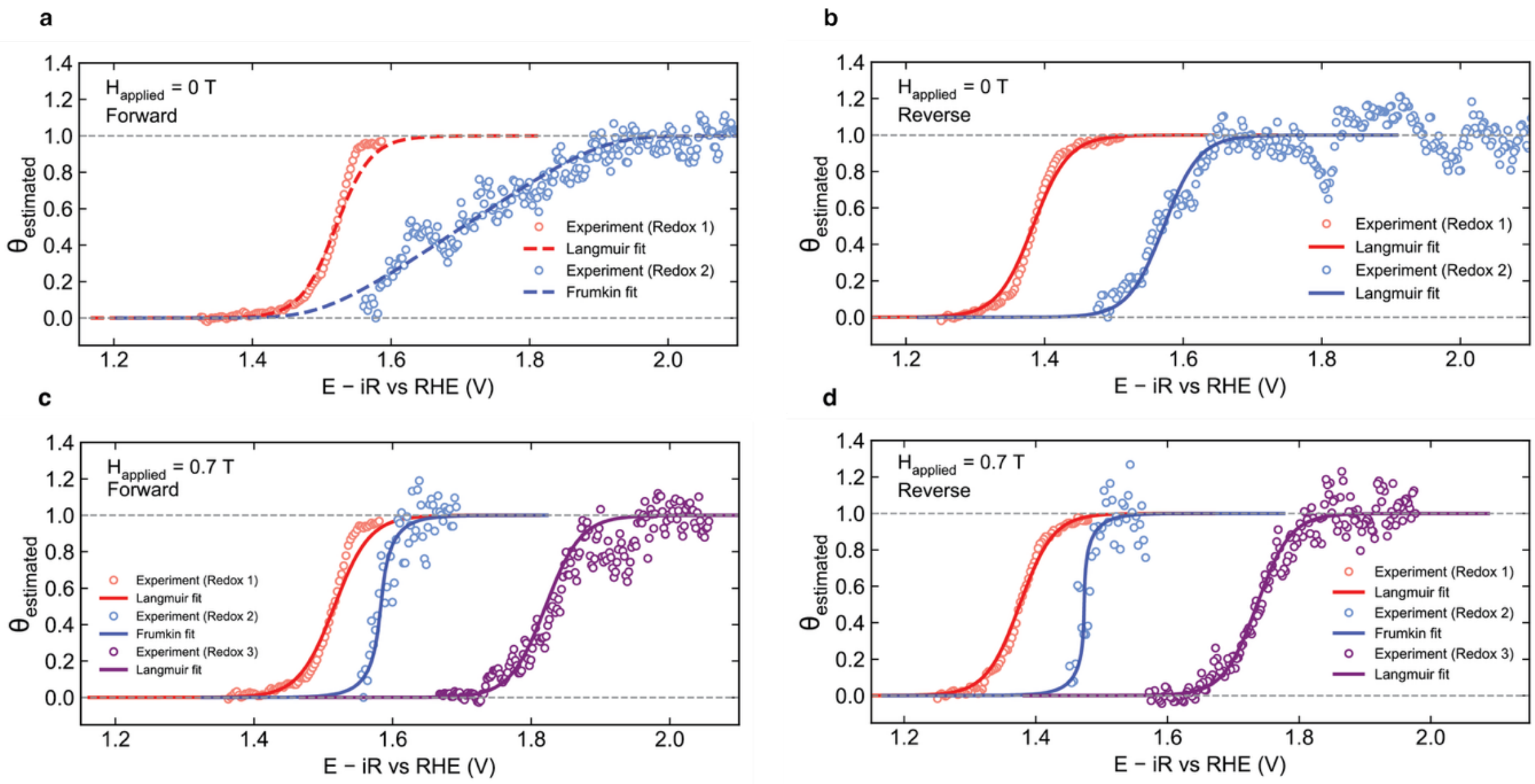

**Figure 5│ Redox transitions as a function of potential. (a-b)** Experimental coverage (θ) of redox transitions as a function of potential without magnetic field in the forward (a) and reverse (b) direction. The Ni(II)/Ni(III) redox transition (red) fits well to the Langmuir electroadsorption isotherm while for the oxo formation step (blue), a good fit is obtained for a Frumkin model with a fitted interaction energy of ~ 0.3 eV for the forward oxidation process. **(c-d)** Experimental coverage of redox transitions as a function of potential under an applied magnetic field of $H$ = 0.7 T in the forward (c) and reverse (d) directions. In this case, the Ni(II)/Ni(III) redox transition (red) exhibits Langmuir-type coverage and oxo formation step (blue) exhibits Frumkin-type coverage. In addition, a new redox transition is observed after the OER onset (purple) which can also be well-described with a Langmuir model. A comparison of the different model outputs is shown Supplementary Note 8.

# Supporting Information for

## *Spin–lattice coupling enables adaptive adsorption in magnetically-driven electrocatalysts*

Arnold Gaje[1†], Lulu Li[2†], Felipe A. Garcés-Pineda[2], Camilo A. Mesa[3*], Ghazaleh Abdolhosseini[2,4], Aditya K. Kushwaha[1], Dora Zalka[5], Elzbieta Trzop[1,7], Nicolas Godin[1], Raffaella Torchio[5], María Escudero-Escribano[3,6], Eric Collet[1,7], Sixto Giménez[8], Niels Keller[1], José Ramón Galán-Mascarós[2,6], Núria López[2*], Ernest Pastor[1,7*]

[1] *CNRS, Univ. Rennes, Institut de Physique de Rennes, Rennes, France.*
[2]*Institute of Chemical Research of Catalonia (ICIQ-CERCA), The Barcelona Institute of Science and Technology (BIST), Av. Països Catalans, 16, 43007 Tarragona, Spain*
[3]*Catalan Institute of Nanoscience and Nanotechnology (ICN2), CSIC, Barcelona Institute of Science and Technology, UAB Campus, 08193, Bellaterra, Barcelona, Spain*
[4]*Departament de Química Física i Inorgànica, Universitat Rovira i Virgili, Marcel. lí Domingo 1, Tarragona 43007, Spain*
[5]*European Synchrotron Radiation Facility, Grenoble 38000, France*
[6]*ICREA, Passeig Lluís Companys 23, 08010 Barcelona, Spain*
[7]*CNRS, Univ Rennes, DYNACOM (Dynamical Control of Materials Laboratory) - IRL2015, The University of Tokyo, Tokyo, Japan.*
[8]*Institute of Advanced Materials (INAM) Universitat Jaume I, 12006, Castelló, Spain*

* camilo.mesa@icn2.cat, nlopez@iciq.es, ernest.pastor@cnrs.fr

† Contributed equally

Contents

## Supplementary Note 1. Theoretical Methodology

DFT simulations were carried out with Vienna *ab initio* simulation package (VASP)[1,2]. We employed Perdew-Becke-Ernzerhof (PBE) density functional to describe the exchange-correlation interactions[3]. We used the Hubbard correction by means of the Dudarev approach[4] with $U_{eff}$ for 5.5 eV of Ni and 5.0 eV of Fe[5] to mitigate electron self-interaction error for the *d* electrons of transition metals. The inner electrons were described with projector augmented-wave (PAW) core potentials[6,7] and valence electrons were represented by plane-waves with a kinetic cut-off energy of 450 eV. The Monkhorst-Pack method[8] was used to generate a Γ-centered mesh with a reciprocal grid size narrower than 0.032 $Å^{-1}$ to sample the Brillouin zone. Spin polarization was included for all calculations considering the magnetic nature of the transition metal systems.

The NiOOH structure was optimized and then used to construct slab models representing the catalytic surfaces. We built (2×2) slab models with 3 atomic layers to represent the NiOOH surface, with a vacuum region of 15 Å added between slabs and a dipole correction[9] included along the z-axis. For the Fe-substituted model, one Ni atom was replaced by Fe in the surface layer to create the NiFeOOH (Fe-substituted) configuration. The Fe adatom model was constructed by placing a Fe atom on the optimized NiOOH surface, following the approach described in previous literature[10]. In all slab calculations, the bottom layer was kept fixed to mimic bulk positions, while the upper layers and adsorbates were allowed to fully relax during geometry optimization.

The convergence criteria for electronic self-consistency and ionic relaxation were set to $10^{-6}$ eV and 0.05 eV/Å, respectively. The conjugate gradient algorithm was employed for ionic relaxation. Gaussian smearing with a width of 0.01 eV was applied for electronic occupations. All calculations included spin polarization and symmetry was turned off to properly account for the magnetic properties and local structural distortions of the Ni and Fe centres. Different magnetic configurations were systematically explored to identify the most stable magnetic states for each system. All DFT structures and computational details can be accessed online in the ioChem-BD repository at https://iochem-bd.iciq.es/browse/review-collection/100/114478/bf71bfb8a255eef85e1ff9b1.

The OER free energy diagrams were obtained using the computational hydrogen electrode (CHE) model[11], and the detailed pathways in two metal-oxyl radical (I2M) mechanism[12] are shown in **Fig. S1**, **S2**, and **S3**.

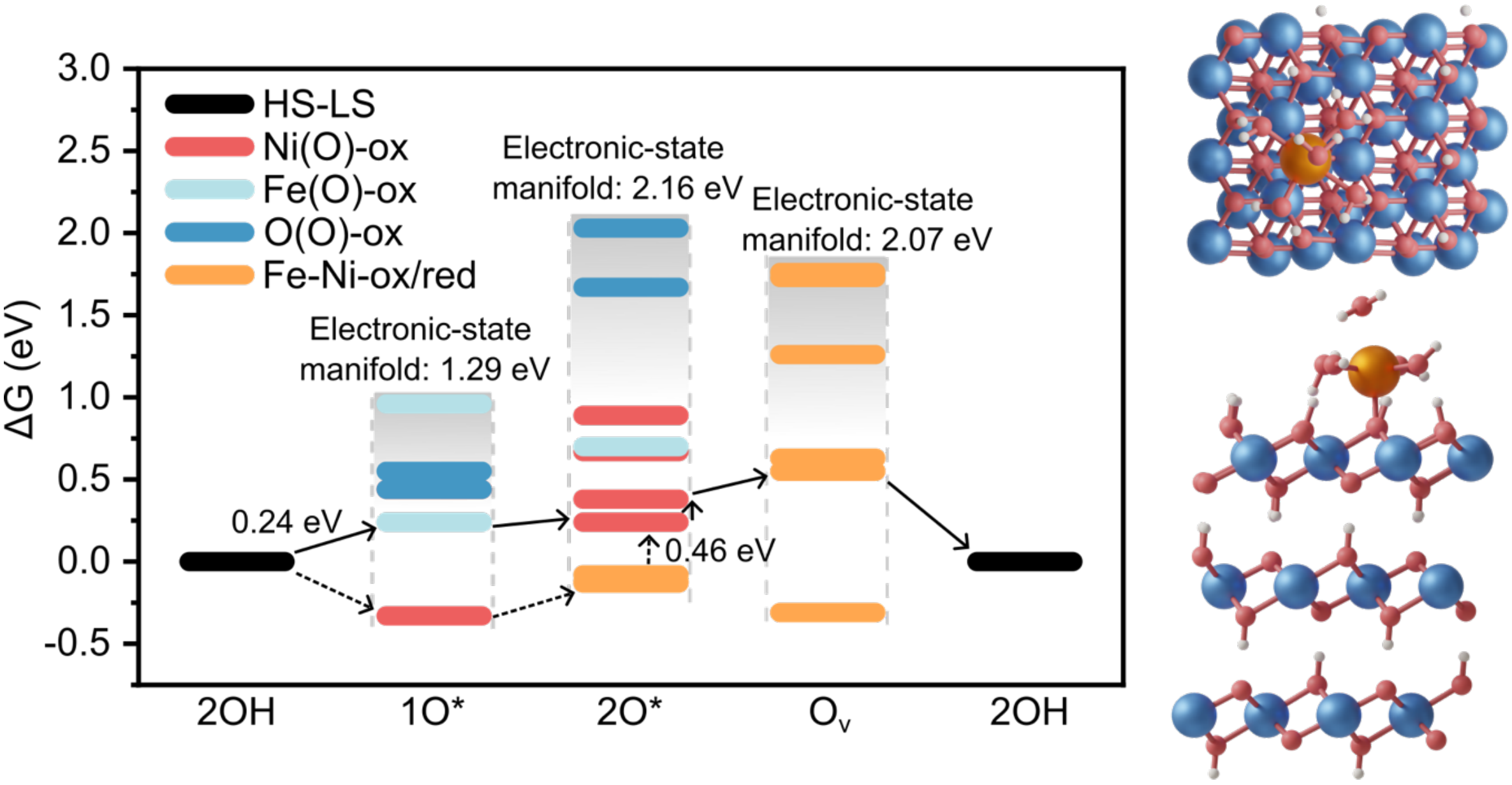

**Figure S1**｜**Free energy pathways and representative structural models for the OER on Fe-adatom NiFeOOH system**. Left: Calculated thermodynamic free-energy landscape of OER intermediates on Fe-adatom NiFeOOH. For each fixed stoichiometry (1O*, 2O*, and Ov), multiple self-consistent electronic solutions are identified, corresponding to distinct oxidation localizations (Ni-, Fe-, O-centered, and Fe-Ni coupled oxidation motifs). These states define electronic-state manifolds spanning 1.29 eV (1O*), 2.16 eV (2O*), and 2.07 eV (Ov), respectively. The solid black line highlights one representative low-energy projection of the manifold along the reaction coordinate. Right: Representative atomic configurations illustrating Fe-adatom–induced oxidation motifs on the NiOOH surface.

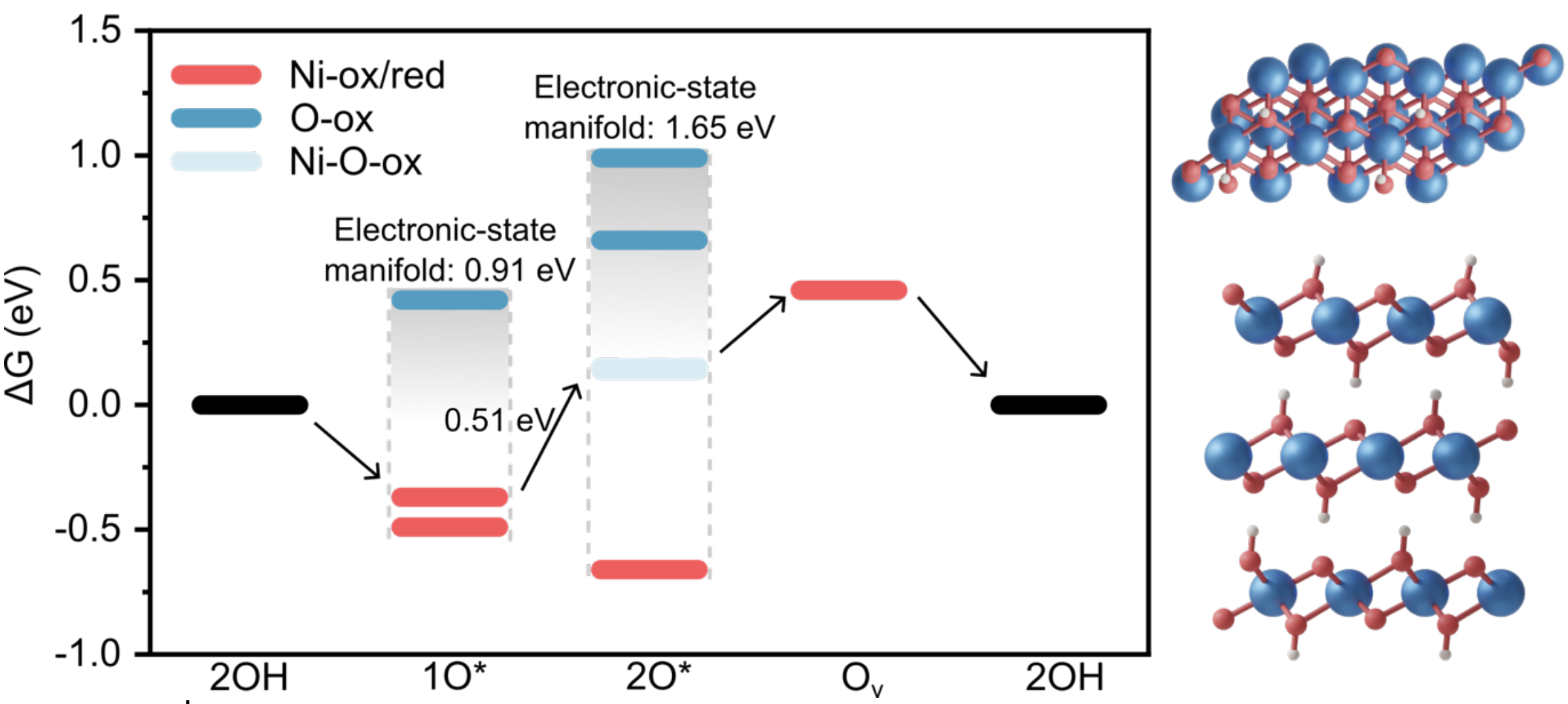

**Figure S2│ Free energy pathways and representative structural models for the OER on NiOOH system.** Left: Calculated thermodynamic free-energy landscape of OER intermediates on NiOOH. For each fixed stoichiometry (1O* and 2O*), multiple self-consistent electronic states are identified, corresponding to distinct oxidation localisations (Ni-centred, O-centred, and Ni-O coupled motifs). These solutions define electronic-state manifolds spanning 0.91 eV (1O*) and 1.65 eV (2O*), respectively, despite identical atomic stoichiometry and surface coverage. The solid black line indicates one representative low-energy projection of the manifold along the reaction coordinate. Right: Representative slab configurations illustrating the distinct oxidation motifs on the NiOOH surface.

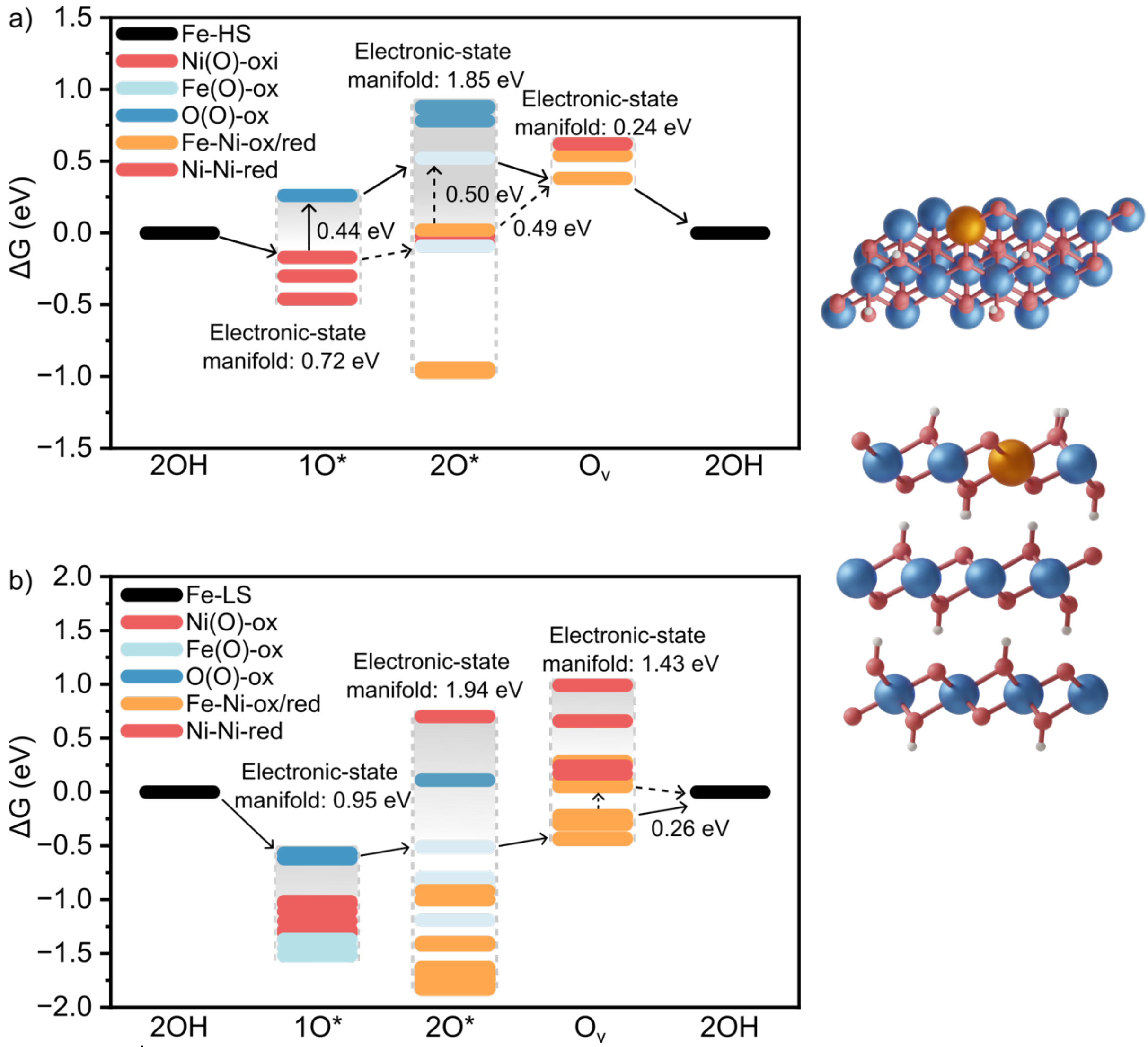

**Figure S3│ Spin-resolved free energy manifolds and representative structures for OER on Fe-doped NiOOH.** (a) High-spin (HS) and (b) low-spin (LS) free energy profiles along the OER pathway on Fe-doped NiOOH. At each intermediate, multiple oxidation localisations (Ni-centred, Fe-centred, O-centred, and Fe-Ni coupled states) are shown, illustrating the electronic-state manifold accessible at fixed stoichiometry and coverage. The solid black line indicates a representative low-energy projection of the manifold along the reaction coordinate. Right panels show representative slab configurations corresponding to distinct oxidation motifs.

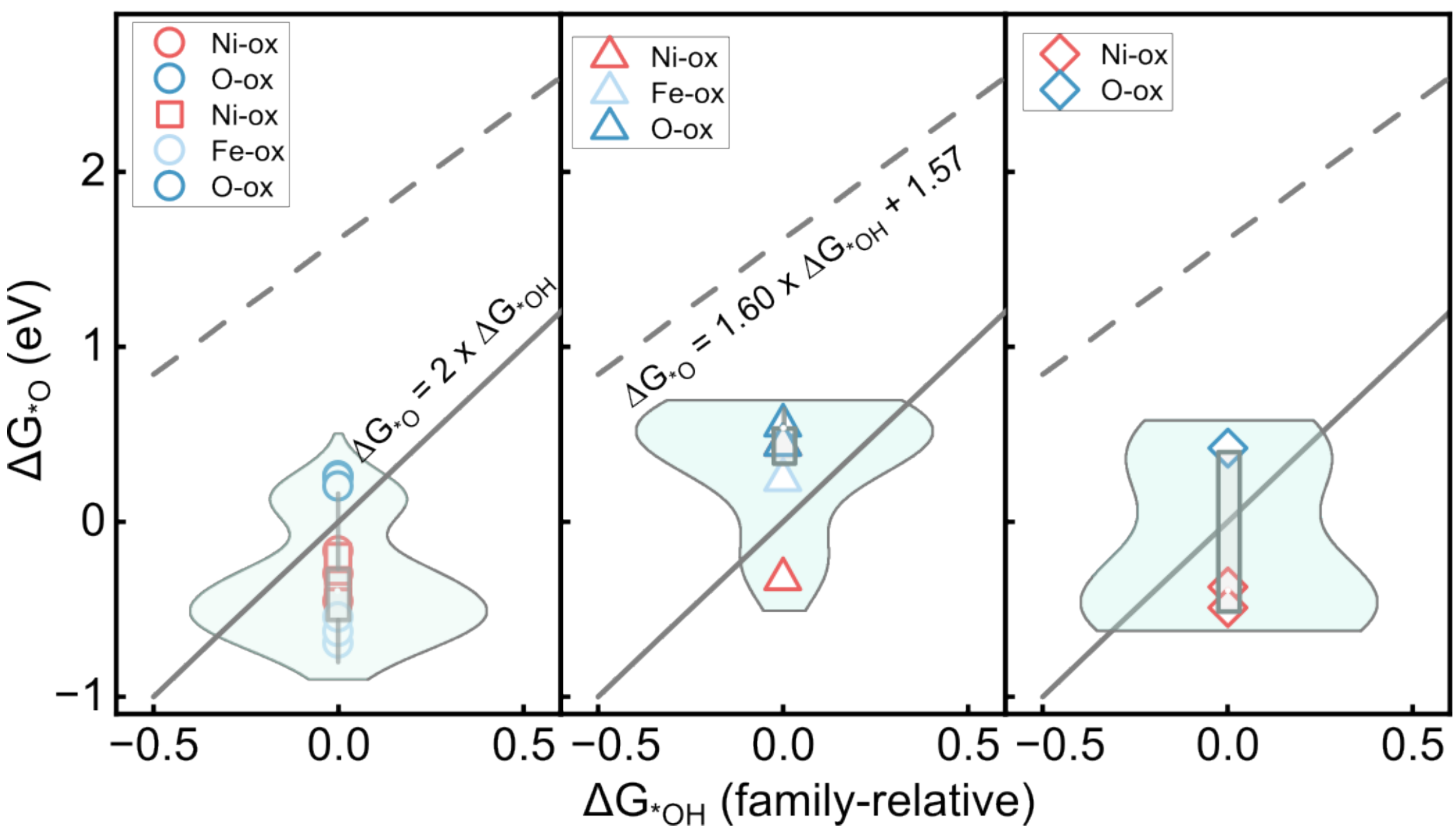

**Figure S4 | Apparent breakdown of the O vs OH adsorption scaling within individual surface families.** Relationship between $\Delta G_{*O}$ and $\Delta G_{*OH}$ for (left) Fe-doped NiOOH, (middle) Fe-adatom NiOOH, and (right) pure NiOOH. For each surface family, $\Delta G_{*OH}$ is referenced to the lowest energy *OH configuration within that family to enable comparison among distinct electronic states at fixed stoichiometry and coverage. Despite nearly identical $\Delta G_{*OH}$ references, $\Delta G_{*O}$ spans a broad energetic manifold, indicating a non-single-valued mapping between the two descriptors. Solid gray lines denote the bond-counting baseline ($\Delta G_{*O} = 2\Delta G_{*OH}$) for literature, while dashed lines indicate the empirical adsorption scaling relation reported in the literature[13]. Violin plots illustrate the distribution of $\Delta G_{*O}$ values arising from different oxidation localizations.

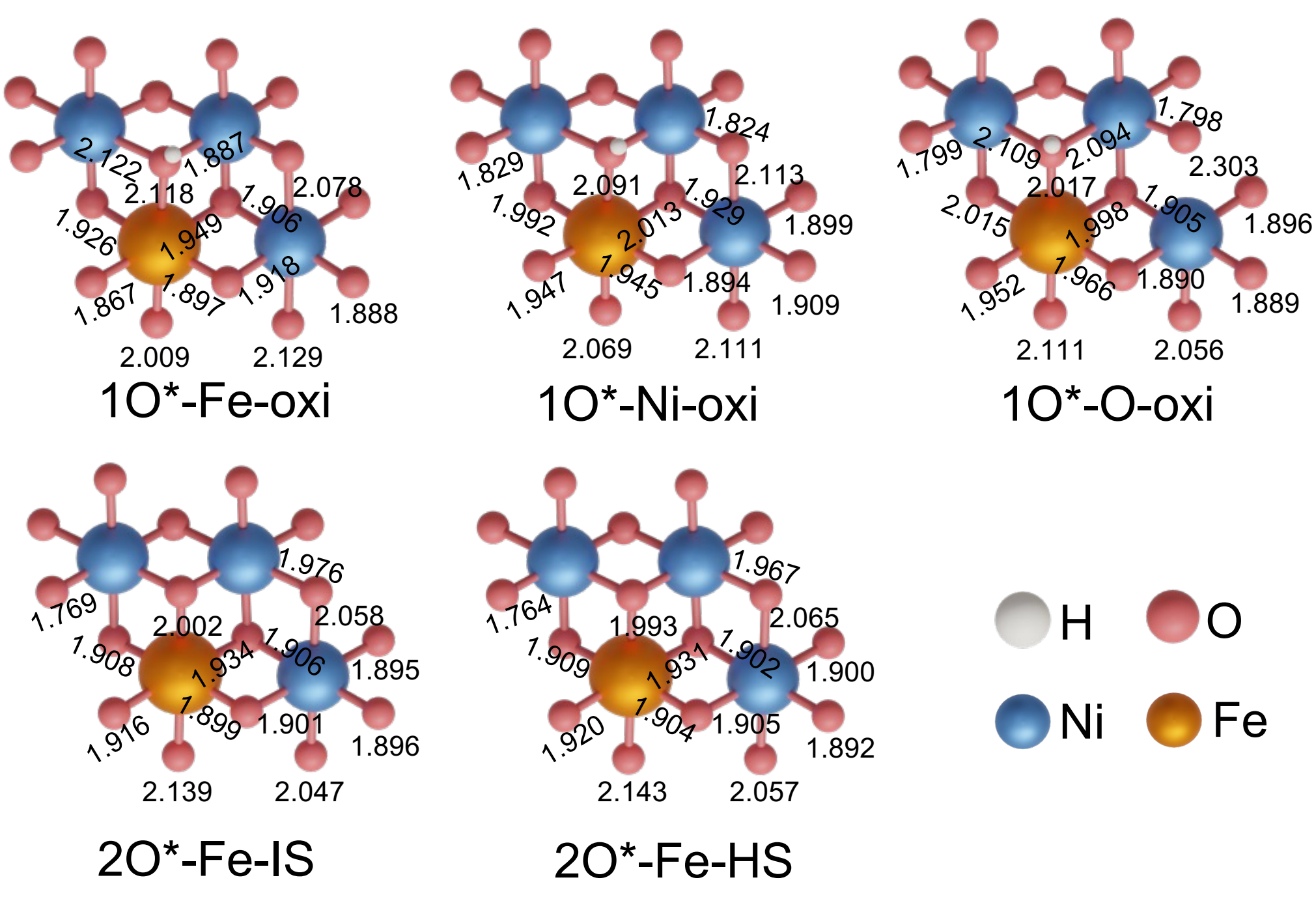

**Figure S5│ Local geometries and M−O bond lengths for representative OER intermediates on NiFeOOH.** Top: Optimized structures for 1O* intermediates showing three oxidation species, Fe-centered (1O*-Fe-ox), Ni-centered (1O*-Ni-ox), and O-centered (1O*-O-ox), highlighting differences in bond lengths associated with oxidation localization. Bottom: Comparison of 2O* Fe-Ni oxidation intermediates with low spin and high spin, highlighting spin-dependent structural distortions. The reported M−O bond lengths are in Å.

## Supplementary Note 2. NiFe oxide synthesis and characterisation

The synthesis and characterisation of the NiFe oxide electrode used in this study has been described in the previous work[14]. In summary, 75 nm NiFe oxide film is deposited on a clean fluorine-doped tin oxide substrate (FTO) using a two-step procedure, involving a physical vapor deposition of $Ni_4Fe$ alloy on FTO and followed by an annealing process in air to oxidise the alloy into a NiFe oxide film. Structural and elemental analyses revealed that the as-synthesised Ni-Fe oxide film is essentially composed of NiO matrix with finely dispersed Ni-Fe spinel ($NiFe_2O_4$) nano islands on its surface, with atomic composition of ~ 82 % Ni and 18 % Fe. The magneto-optical measurements described in Supplementary Note 3 show that this sample exhibits magnetic properties that are characteristic of magnetically-active mixed Ni-Fe oxide thin films[15–18].

This NiFe oxide shows OER electrocatalytic activity, typical of Fe-doped Ni oxides[19–21]. As demonstrated here (see **Fig. 3a** and **3b** in the main text) and in previous works[14,22], it is also a magnetically active catalyst, exhibiting an OER activity that is enhanced by the application of an external magnetic field.

## Supplementary Note 3. Magneto-optical measurements

The measurement of static magneto-optical (MO) properties of the as-synthesized NiFe oxide electrode was performed using a custom-designed broadband magnetic circular dichroism spectrometer based on a 90°-polarisation modulation technique. Details of the experimental set-up are described in references[23–25]. The MO properties were measured via the Faraday effect in a polar configuration, where the magnetic field and the incident light were applied perpendicular to the sample. A continuous white light emitted by a 150 W Xe arc lamp was used as a probe. The emitted light was polarised by a Glan-Taylor prism and modulated at a high frequency of 50 kHz by a Hinds photoelectric modulator (PEM). The modulated light is focused onto the sample at normal incidence. After interaction with the sample, the transmitted beam was collimated into a polarisation analyser and then focused into a monochromator to select the desired probe wavelength. The final intensity of light is transformed into an electric current by a photomultiplier detector and demodulated by two lock-in amplifiers referenced to the first and the second harmonics of the PEM. The signal referenced to the first harmonic is proportional to the Faraday ellipticity ($\varepsilon_F$), whereas the one referenced to the second harmonic is proportional to the Faraday rotation ($\Theta_F$).

The spectral dependency of the Faraday signal $X_F$ (with X = Θ or ε) is obtained from the difference between the $X_F$ spectra measured for positive and negative saturating external magnetic fields of $H_{ext,sat}^+ = +0.8$ T and $H_{ext,sat}^- = -0.8$ T, using the following equation: $X_F = [X(H_{ext,sat}^+) - X(H_{ext,sat}^-)]/2$. This avoids any contribution that is not proportional to the magnetisation in the MO spectra[25,26].

To obtain the magnetic hysteresis loops, the MO $\varepsilon_F$ signal at λ = 542 nm incident light was measured at room temperature in a quasi-static mode with a field sweep rate of 4 mT $s^{-1}$. The small Faraday contribution induced by the paramagnetic substrate was carefully subtracted from the total hysteresis loops to obtain the intrinsic Faraday signal of the film. Then, a normalised magnetic hysteresis loop $M/M_s$ (**Fig. 3c** in the main text) is obtained using the following equation $M/M_s = X_F(H_{ext})/X_F(H_{ext,sat})$.

The optical and magneto-optical measurements were performed in the photon energy range of 1.5 to 3.5 eV. The optical and magneto-optical spectra of the film show features characteristic of single-crystalline ferrimagnetic $NiFe_2O_4$ [27] (**Fig. S6b**). In the studied spectral range, we obtain satisfactory agreement of the optical absorbance (OD) and Faraday ellipticity spectra (**Fig. S6a-b**) with experimental data observed previously in Ni-Fe oxide samples[27–29]. The magneto-optic hysteresis loops presented in **Fig. 3c** of the main text also show the typical magnetisation behaviour of Ni-Fe oxide mixed systems. Specifically, the coercive, saturation, and exchange bias fields from the magneto-optic hysteresis loops are closely similar to previous experimental results for pure and mixed Ni-Fe oxide systems[15–17,27].

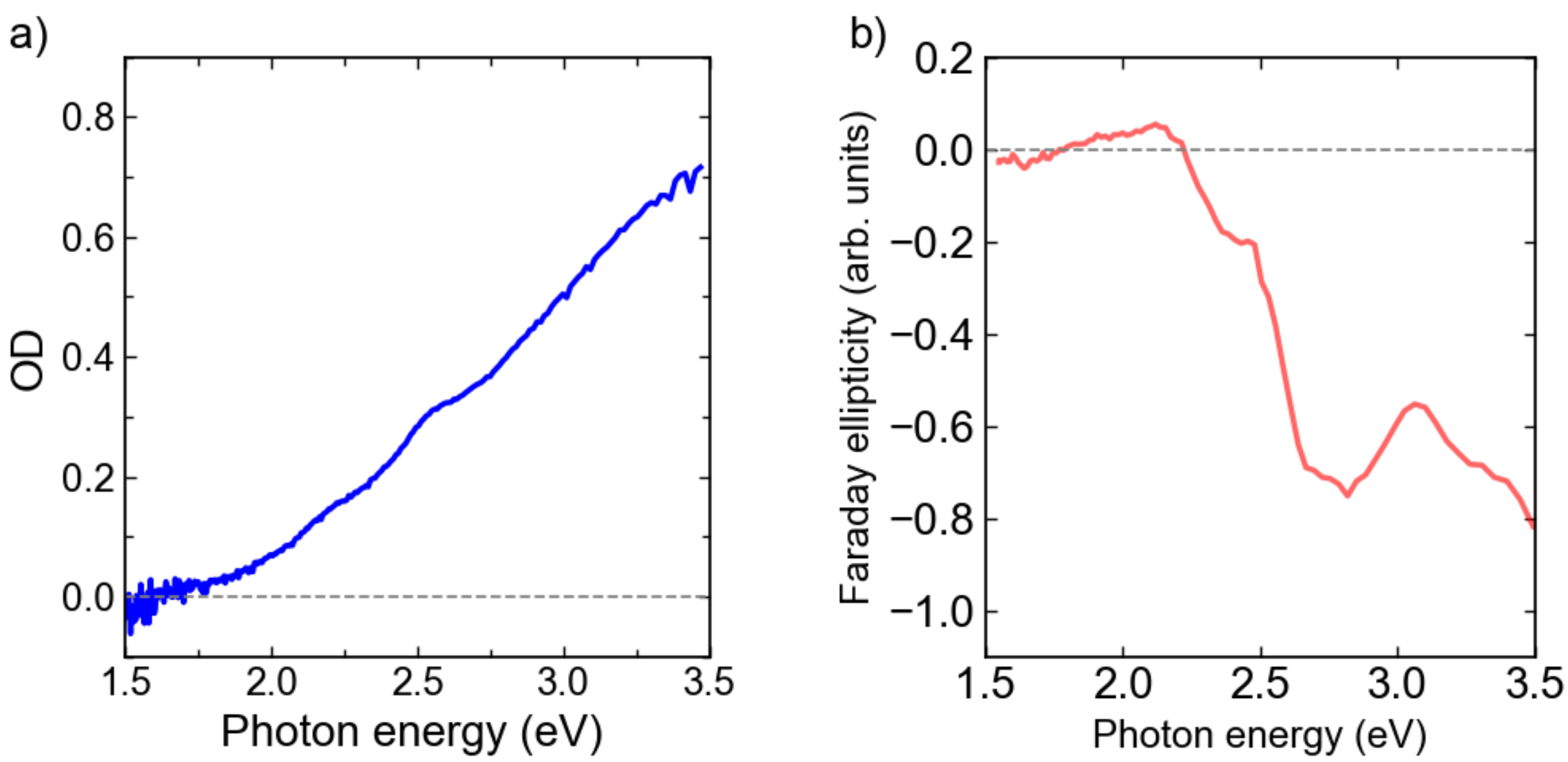

**Figure S6│ Optical and magneto-optical spectra of the nickel-iron oxide system.** (a) Optical absorption spectrum and (b) Faraday ellipticity spectrum measured on pristine 75 nm thin film deposited on 1.1 mm thick FTO-coated conductive glass substrate.

## Supplementary Note 4. X-ray Absorption Spectroscopy

To complement the spectroelectrochemical characterisation of the catalytic material, we performed ex-situ X-ray absorption spectroscopy (XAS) on the sample after catalyst activation. When the catalyst activation procedure (see Supplementary Note 5a) was finished, the electrode was subsequently removed from the cell and placed in a custom-designed sample holder for ex-situ XAS measurements in fluorescence mode. X-ray absorption spectroscopy (XAS) measurements were performed at the BM23 beamline of the European Synchrotron Radiation Facility (ESRF, Grenoble, France). Data were collected at the Fe K-edge (7.112 keV) and Ni K-edge (8.333 keV). The incident X-ray energy was selected using a Si(111) double-crystal monochromator (DCM). Harmonic rejection was achieved using the BM23 double-mirror system. For the Fe and Ni K-edge measurements, the mirrors were operated on the Si stripe at an incidence angle of 3 mrad.

The incident and transmitted intensities – used for the reference samples – were recorded using three ionization chambers (I0, I1, and I2), each with an effective length of 30 cm and operated at 1 kV. The incident ionization chamber (I0) was filled with 0.989 bar $N_2$, complemented with He to a total pressure of 2 bar, while the transmission chambers (I1 and I2) were filled with 0.221 bar Ar and similarly topped up with He to 2 bar. The energy calibration of the monochromator was verified by measuring reference spectra of metallic Fe and Ni foils. All spectra have been recorded in continuous mode.

For fluorescence measurements of the NiFe oxide samples, the sample geometry was set to 45 degrees with respect to incident X-ray beam and fluorescence was collected using a 2-mm Parker Si SDD Vortex single element placed at 90 degrees with respect to X-ray beam.

Fluorescence signal from Fe and Ni was isolated from the total fluorescence by selecting Fe and Ni ROIs (Regions of Interest), calibrated using reference Fe and Ni foils.

Figure S7 displays the Fe and Ni K-edge XANES and EXAFS spectra from these measurements. The XAS spectra of reference compounds collected in transmission mode are also shown for comparison. The XANES and EXAFS spectra for the unactivated pristine samples are shown in red. The measured response is in good agreement with the response of as-prepared NiFe mixed oxides [30–35]. At the Fe K-edge (**Fig. S7a**), little change in the XANES spectra can be observed before and after activation, indicating that electrochemical activation does not significantly alter the oxidation state of Fe. This is in agreement with the results of previous studies on NiFe oxides OER catalysts[30–35]. The EXAFS spectra reveal a slight decrease in the average Fe-O bond distances and the broadening of their distribution (**Fig. S7c**). At the Ni K-edge spectra (**Fig. S7b**), there is also no substantial shift in the pre-edge peaks and edge position, indicating that the average oxidation state of Ni in the sample remains the same after activation. We observe a small increase and broadening in the white line, pointing towards small changes in the bonding structure and coordination environment.

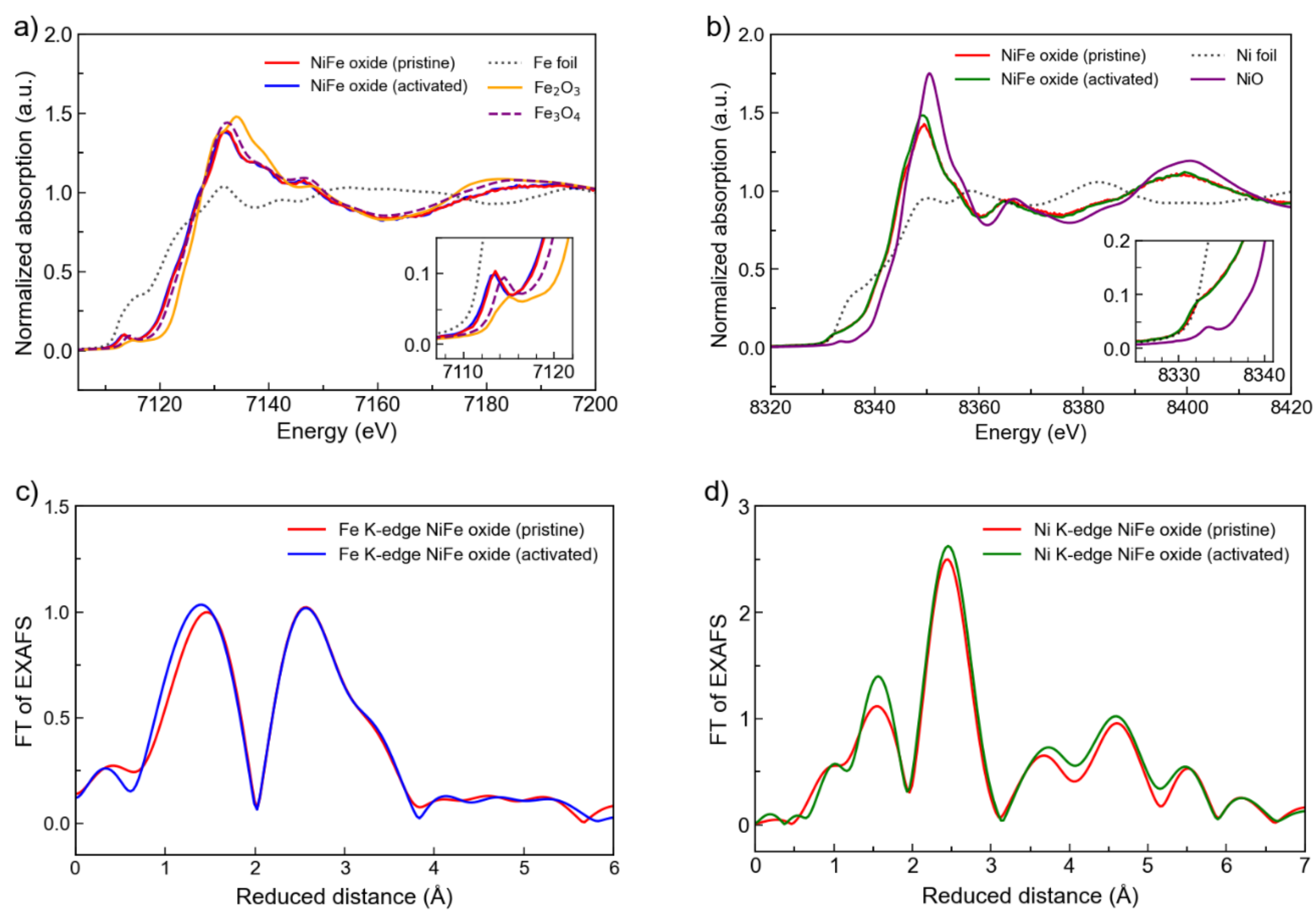

**Figure S7. Ex-situ Ni and Fe K edge XAS spectra for the activated NiFe oxide catalyst. a-b** Normalised XANES data of (a) Fe and (b) Ni K-edge. Features of the pre-edge are shown in the inset. **c-d** Fourier transformed EXAFS spectra of (c) Fe and (d) Ni K-edge. XAS spectra of reference compounds are also shown for comparison.

## Supplementary Note 5. *In-situ* spectroelectrochemistry measurements

All the spectro-electrochemical experiments were conducted by fitting a spectroelectrochemical cell in a custom-designed magneto-optics set-up described in the Supplementary Note 3. This set-up, schematically shown in **Fig. 4a** of the main text, allowed us to measure the optical and electrochemical response of the sample as a function of applied potential at different external magnetic fields.

### a. Catalyst activation and electrochemical response

Before the spectroelectrochemistry experiments, a pre-conditioning step was performed on all the electrodes to reach consistent, repeatable and stable electrocatalytic behaviour. This procedure is similar to that typically applied to Ni-based electrodes and involves 300 to 400 successive cyclic voltammetry (CV) cycles at a scan rate of 75 mV $s^{-1}$ to form the OER-active Ni-Fe hydroxide/oxyhydroxide (**Fig. S8a**). To check that stable electrocatalytic behaviour is achieved after activation, 2 to 3 CV were measured at a slow scan rate of 10 mV $s^{-1}$ at H = 0 T and H = 0.7 T.

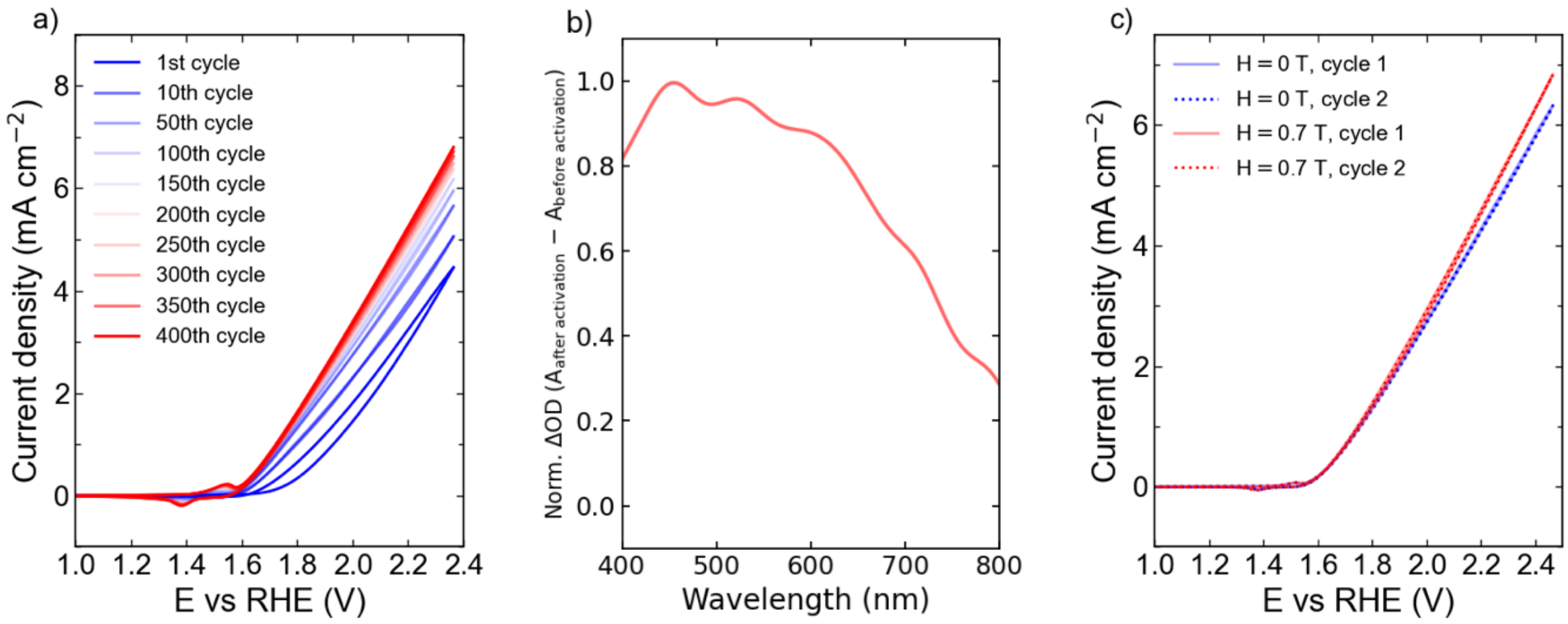

**Figure S8│ Activation protocol**. (a) 400 CV curves obtained at a fast scan rate of 75 mV $s^{-1}$ in a typical activation protocol. (b) Normalised ΔOD spectrum obtained from the difference between the absorbance spectra at open circuit potential (OCP) after and before activation. For visualization of the main features of the spectrum, the ΔOD data are smoothed by Gaussian filtering. (c) 2 Consecutive CV curves of an activated sample at H = 0 T and H = 0.7 T obtained at a scan rate of 10 mV $s^{-1}$. The difference between consecutive measurements is negligible, indicating the stable electrochemical response of the activated electrode.

The CV of the catalyst after activation shows the characteristic features of NiFe oxyhydroxides: a positively shifted $Ni^{2+}/Ni^{3+}$ redox potential peak and a decreased OER onset potential[20,36–39]. The UV-Vis spectrum (**Fig. S8b**) was obtained by subtracting the initial absorbance of the sample from the absorbance after the activation at open-circuit potential. It exhibits a broad absorption feature below 800 nm with prominent peaks at ca. 450 nm and 530 nm in agreement with previous measurements of NiFeOOH[40]. Both experimental results confirm the growth of NiFe oxyhydroxide through the activation process described above. Furthermore, stable electrochemical response has been obtained after this activation step (**Fig. S8c**).

Comparison of the simultaneously measured CV and optical signals (**Fig. S9**) at different stages during the experiments shows that the sample exhibits stable electrochemical behaviour after activation, and during the *operando* measurements at the selected applied potential range. As shown in in **Fig. S9**, we found the reversible response of the CV and optical signals (at both H = 0 T and H = 0.7 T) in the voltage range measured (note the analysis in the main text focuses on a smaller voltage range where optical saturation regimes could be identified as described in Supplementary Notes 7 and 8) further indicating that the systems remains active during the course of our operando or in-situ spectroelectrochemistry measurements.

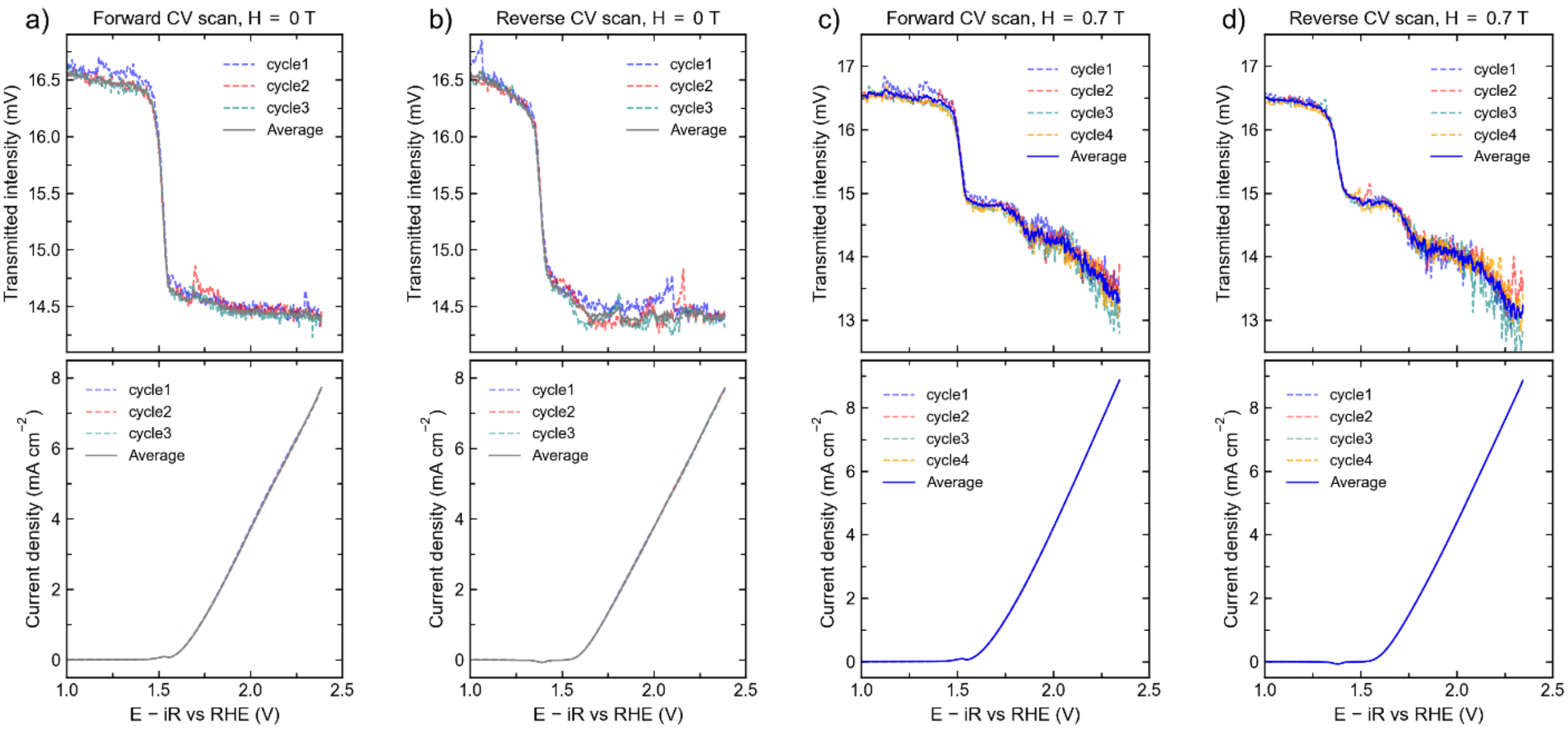

**Figure S9**| **Cyclic voltametric response.** Current density and optical signals at (a, b) H = 0 T and (c, d) H = 0.7 T during multiple CV cycles. In this in-situ optical spectroscopy experiment, the transmitted light intensity and current density are simultaneously recorded during the forward and reverse scan of each CV cycle in a 3- or 4-CV measurement. The measurements are done using a 10 mV $s^{-1}$ scan rate.

**b. Electrochemical measurements.** The electrochemical measurements were performed under ambient conditions using a Biologic SP-150e two-channel, EIS-capable potentiostat, in a three-electrode configuration with a coiled Pt wire counter electrode and an Ag/AgCl (saturated KCl) reference electrode. The electrolyte solution consisted of 0.1 M KOH (pH 13). All recorded potentials were referenced against the Ag/AgCl electrode and converted to the reversible hydrogen electrode (RHE) reference scale using the formula $E_{RHE} = E_{Ag/AgCl} + 0.197 + 0.059$ pH (V). Current densities J were calculated based on the geometrical surface area of the electrodes. EIS (ZIR) was used to estimate the Ohmic drop that was found to be around 24 Ohms, in agreement with previous measurements of the same type of electrodes[14]. The correction was applied during the data analysis step and not during the measurements. For magnetic field dependence electrochemical measurements (**Fig. 3b** in the main text and **Fig. S10**), the data were collected at a fixed potential while switching the applied magnetic field from + 0.9 T to – 0.9 T at a field sweep rate of 4 mT $s^{-1}$ that allows quasi-static measurement.

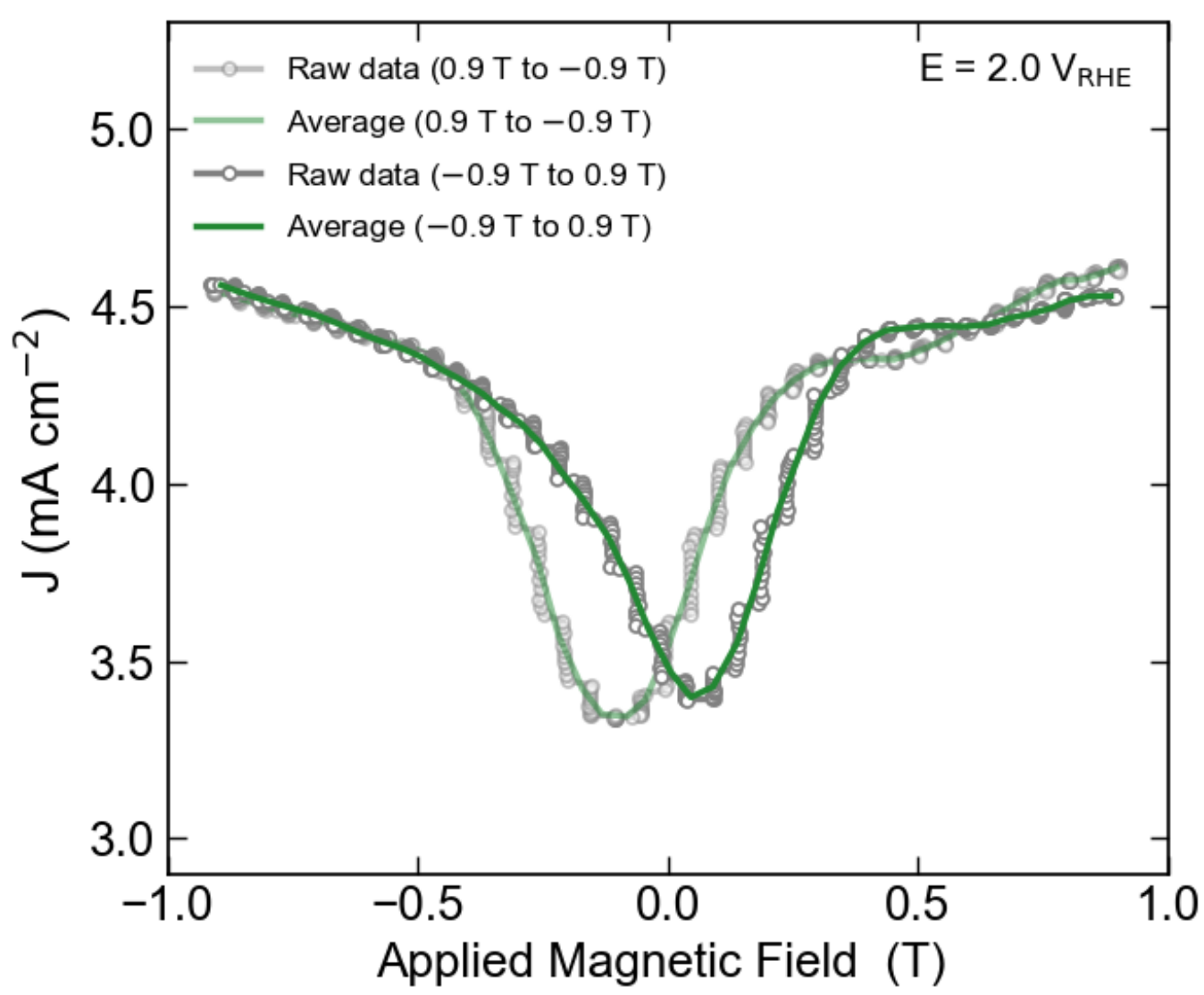

**Figure S10 | Magnetic field dependence of OER activity.** The current density at 2.0 $V_{RHE}$ (at which magnetic enhancement can be observed) was measured while switching the applied magnetic field from + 0.9 T to – 0.9 T. This measurement was done in a quasi-static mode with a scan rate 4 mT $s^{-1}$. Here, the magnetic field was ramped with a 0.04 T step, and for each field step the current was recorded every 1 s for 10 s. Thus, at a particular field, 10 measurements of current were obtained (plotted points/lines in gray). The average of these 10 measurements was obtained and plotted against magnetic field (plotted lines in green as shown in the main text).

**c. Optical measurements.** In-situ UV–visible absorption spectroscopy measurements were conducted in transmission mode. A collimated white light emitted by the 150 W Xe arc lamp light source was transmitted through the sample inside a custom-built electrochemical cell. The light was then collected and recollimated using planoconvex lenses and refocused into a CCD-based spectrometer (Ocean Optics TR2000). The spectra were collected between 350 and 800 nm with an integration time of 250 ms. Potentials were controlled using a Biologic SP-150e Potentiostat (BioLogic, Seyssinet-Pariset, France). The measurements were made under potentiostatic conditions using the same electrochemical set-up mentioned above, and the spectra were collected after the current density reached a steady state. The measured data are reported as spectroelectrochemical difference spectra (ΔOD), which are generated by subtracting a reference spectrum (in this case, the OD at 1.25 $V_{RHE}$, at which the catalyst is in a resting state) from the absorption data obtained under the different conditions of interest[41].

In addition to steady-state measurements, spectroelectrochemistry was also performed by scanning the applied potential at a set rate of 10 $mVs^{-1}$. For these measurements, the stabilised white light emitted by the 150 W Xe arc lamp light source was modulated by a chopper at a frequency of 205 Hz, providing a synchronous reference signal for lock-in detection. The modulated light was then focused onto the sample at normal incidence. After interaction with the sample, the transmitted light was collimated and sent to a monochromator set at a selected wavelength of 542 nm. This wavelength was selected as it had previously been used to track the changes in the oxidation state of Ni and Fe centres in Ni(Fe)-based OER electrodes (see Supplementary Note 6 for spectral dependence). The final intensity of monochromatic light was detected by a photomultiplier equipped with a current-to-voltage amplifier, and the resulting signal was determined using a lock-in amplifier. The signal of several scans was averaged in order to improve the signal-to-noise ratio.

The differential absorbance (ΔOD) was calculated with respect to the starting potential of the CV scan as:

$$\Delta\mathrm{OD} = -\log\left(\frac{I_V}{I_{V_o}}\right)$$

where $I_V$ is the intensity of transmitted light at a particular measured potential and $I_{V_o}$ is the intensity of transmitted light at the potential at the reference potential (in this case at E = 1.0 $V_{RHE}$).

## Supplementary Note 6. Tracking the redox transitions using in-situ UV-Vis absorption spectroscopy

We probe the redox chemistry in the NiFe-based electrode in the absence (H = 0 T) or presence (H = 0.7 T) of an applied magnetic field using in-situ UV-Vis spectroscopy (Supplementary Note 5c). Specifically, we record the change in the UV-Vis absorption spectra (relative to 1.25 $V_{RHE}$) when the potential is increased in 50 or 100 mV steps from 1.25 V to 2.1 V. Such spectroscopic measurements have previously been used to track the surface species of electrocatalysts as a function of potential[20,40–46].

At H = 0 T, the spectral changes with increasing potential (**Fig. S11a**) reflect the redox transitions in NiFe oxyhydroxide systems, which have been previously evaluated in several studies[20,38,40,47–49]. Several key features can be observed. First, a sharp increase in ΔOD between 1.4 V to 1.6 V that coincides with the redox peak in the CV before the onset of OER. These spectra of the oxidised states have a peak centred at ca. 500 nm, which has been attributed to the $Ni^{2+}/Ni^{3+}$ redox transition, particularly the oxidation of $Ni(OH)_2$ to NiOOH ($Ni(OH)_2 + OH^- \rightarrow NiOOH + H_2O + e^-$)[20,37]. The associated increase in the optical absorption has been assigned to the nickel *d*−*d* interband transition[47].

Upon increasing the potential to the OER regime, the ΔOD gradually increases from 1.6 V to 2.0 V and tends to saturate at higher potentials. These changes in OD suggest the further oxidation of Ni centres, and this redox process has been attributed to the formation of metal-oxo species: $NiOOH + OH^- \rightarrow NiOO^* + H_2O + e^-$ [20,38,50–52]. For Fe-incorporated NiOOH, the increased absorption at longer wavelength may also suggest that both Ni and Fe metal centres can accumulate oxidation equivalents[40,53], which is consistent with the results of DFT calculations reported in this current study.

In the presence of an applied magnetic field (H = 0.7 T), we observe a further change in OD from ~ 1.8 V to 2.2 V (**Fig. S11b**), which corresponds to the potentials where a significant increase in current density is observed relative to electrochemical measurements at H = 0 T (**Fig. 3a** in the main text). The differential spectra for this new redox transition show a further broadening of the absorption from 500 nm to 800 nm. These observations suggest that the magnetic perturbation triggers changes in the electronic state of the catalyst, which manifest in colouration changes, and which correlate with the current-voltage response of the electrode.

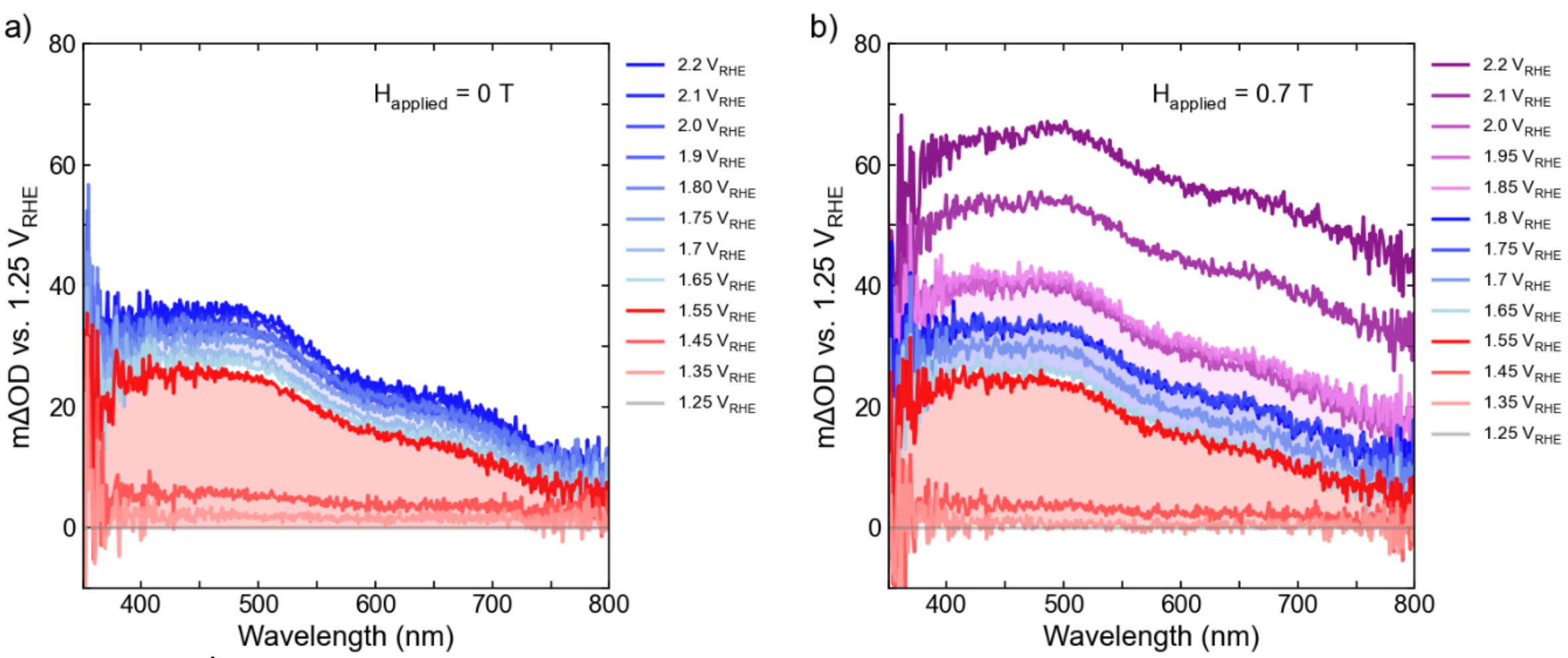

**Figure S11│ Optical changes as a function of applied potential and magnetic field stimulation.** Spectroelectrochemical differential absorbance spectra of the electrode relative to 1.25 $V_{RHE}$ at (a) H = 0 T and (b) H = 0.7 T.

## Supplementary Note 7. Estimation of surface coverage of the redox intermediates

We also performed in-situ UV-Vis spectroelectrochemical measurements with lock-in detection at H=0 T and H=0.7 T to establish whether the presence of a magnetic perturbation induced changes in the reactive intermediates and oxidation degree of the electrode. Single-wavelength lock-in detection was necessary to achieve low noise levels and capture the effects of the external perturbation as a function of the driving field with high voltage resolution. This data collection strategy, which we apply to spectro-electrochemistry, is commonly used in time-resolved spectroscopic experiments, where small signals that are over-imposed on a large noise background need to be detected[54–57].

As shown in **Fig. 4c-f** of the main text, the optical signals exhibit a complex response to the applied potentials. Notably, the response is characterised by several slope changes, which we attribute to the formation of different reactive intermediates (see also discussion in Supplementary Note 6). In this section, we discuss the procedure we applied to isolate such species to establish how their population (or surface coverage) changes as a function of the applied potential. Our goal is to estimate the applied potential at which the concentration of an intermediate species starts to increase and the potential at which it reaches saturation, as these can be approximated to the potentials of 0 and full coverage of its redox transition. We assessed the signal using two analysis methods and estimated the voltage ranges where the different species form. These methods are based on the direct inspection of the $\Delta$OD signal (Method 1) and the derivatives of the $\Delta$OD (Method 2). We obtain a good agreement between the two approaches and conclude that at least 2 intermediates (3 in the presence of the magnetic field) are present. We note that UV-Vis spectroscopy lacks the high chemical sensitivity of other techniques, such as XPS, and consequently, more species could be present. Despite its limitations, our experimental approach allows us to systematically assess and compare with a high voltage resolution the optical changes in the electrode with and without a magnetic field. Such a resolution is needed to contrast the experimental data with established models of surface coverage and extract thermodynamic parameters (Supplementary Note 8) as demonstrated in previous works[42–45].

**Method 1: Visual examination of the slopes in the optical signal as a function of potential. Fig. S12a** shows the optical signal−voltage response in the absence of the magnetic field for the forward redox or oxidation process. We observe several slope changes, around 1.47 V, 1.56 V and 1.90 V vs RHE. The first slope inflexion manifests as a large and abrupt change in OD and aligns with the redox wave associated with the $Ni^{2+}/Ni^{3+}$ transition (see the CV curves in **Fig. 4c**, **Fig. S9a and Fig. S9c**). Upon further increasing the potential above ~ 1.56 $V_{RHE}$, a second change in slope occurs, which is characterised by a more gradual change in OD up to 1.90 $V_{RHE}$, where the signal tends to saturation. This second OD change occurs in the region where OER takes place and has been previously assigned to the formation of oxo species, as discussed in Supplementary Note 6. Based on the intercept between the slopes and the attainment of signal saturation, we estimate that the first redox transition takes place between 1.47 V and 1.56 V vs RHE while the second transition takes place between 1.56 V and 1.90 V vs RHE. **Fig. S12c** displays the $\Delta$OD−potential response of the sample during the reverse scan in which similar changes are observed.

Upon the application of a magnetic field, a similar response can be observed (**Fig. S12b** and **S12d**), although with some key differences (i), the second forward redox process after 1.55 V unfolds over a much narrower voltage range and a saturation is observed at approximately 1.62 to 1.7 $V_{RHE}$ and (ii) a third slope change occurs around 1.75 $V_{RHE}$ which is followed by a signal that tends to saturation at ~1.90 to 2.1 $V_{RHE}$. Based on the intercept between the slopes and the attainment of signal saturation, we estimate that the first redox transition takes place between 1.48 $V_{RHE}$ and 1.55 $V_{RHE}$, the second transition occurs between 1.55 $V_{RHE}$ and 1.63 $V_{RHE}$ and the third transition between 1.76 $V_{RHE}$ and 1.90 $V_{RHE}$.

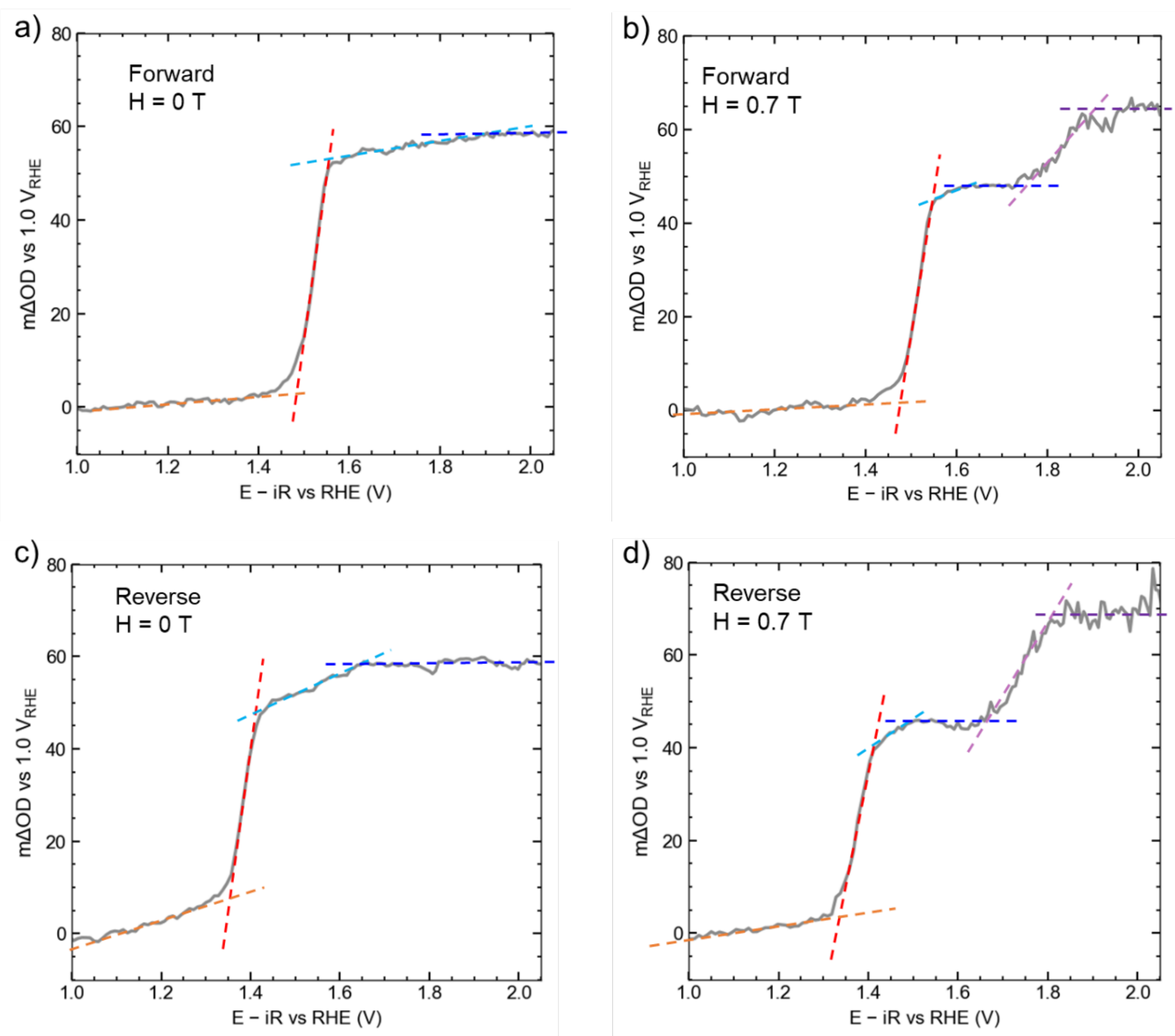

**Figure S12│ Isolation of redox transitions.** Identifying the 0 and full coverage of redox transitions using the different slopes in optical signal as a function of potential at (a,c) H = 0 T and (b,d) H = 0.7 T. This analysis is done for the (a,b) forward/oxidation and (c,d) reverse/reduction scans of the CV. The data sets analysed are average of measurements from 3 or 4 CV cycles.

**Method 2: Derivative of the optical signal as a function of potential**. The derivatives of the optical signal over the potential (E) can more precisely indicate the inflexion points and therefore the applied potentials associated with each redox transition[43]. We obtain both the $1^{st}$ and $2^{nd}$ derivatives of the ΔOD vs E data to determine the voltages at which a transition starts and ends, and which we approximate to the voltages at which the surface coverage is 0 and 1, respectively. The results at H = 0 T and H = 0.7 T are shown in **Fig. S13** and **S14**. We approximate the onset of the first redox transition to the applied potential at which the $1^{st}$ derivative of the optical signal starts to increase, and the full coverage (which is also assigned to the onset of the second redox transition) when the $1^{st}$ derivative becomes zero (in the case of H = 0T) and minimum (in the case of H = 0.7 T). The latter is indicative that the sharp signal rise corresponding to $1^{st}$ redox transition is over. We assign the full coverage of the second redox transition to the point where the optical signal starts to saturate (blue dashed line). At H = 0 T, this occurs at ~ 1.9 $V_{RHE,}$ while at H = 0.7 T this point occurs at a lower potential (~ 1.63 $V_{RHE}$). Similarly, at H = 0.7 T, we assign the full coverage of the third redox transition at the voltage where the optical signal saturates (~ 1.89 $V_{RHE}$). Based on the derivatives of the optical signals and the attainment of signal saturation, we estimate the following ranges: first redox transition, 1.44 $V_{RHE}$ to 1.57 $V_{RHE}$; second redox transition, 1.57 $V_{RHE}$ to 1.90 $V_{RHE}$ for H = 0T and 1.57 $V_{RHE}$ to 1.63 $V_{RHE}$ for H = 0.7 T; and third redox transition, 1.72 $V_{RHE}$ to 1.89 $V_{RHE}$ (for H = 0.7 T only).

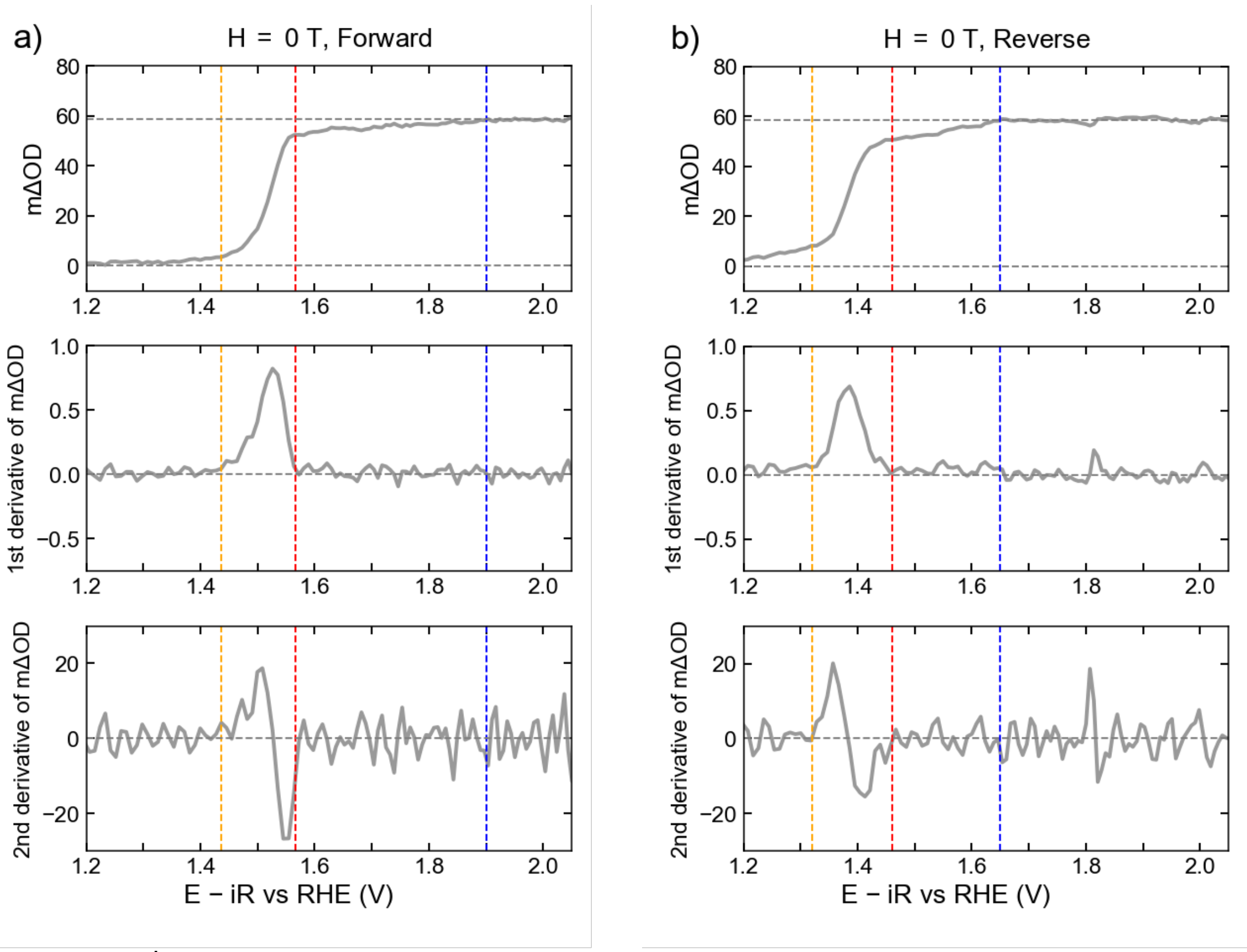

**Figure S13│ Isolation of redox transitions with the derivative method.** Identifying the 0 and full coverage of redox transitions at H = 0 T using the derivative of the optical signal as a function of the potential for the forward or oxidation process (a) and reverse or reduction process (b). Dashed vertical lines indicate the potentials for 0 and full coverage of the redox transitions.

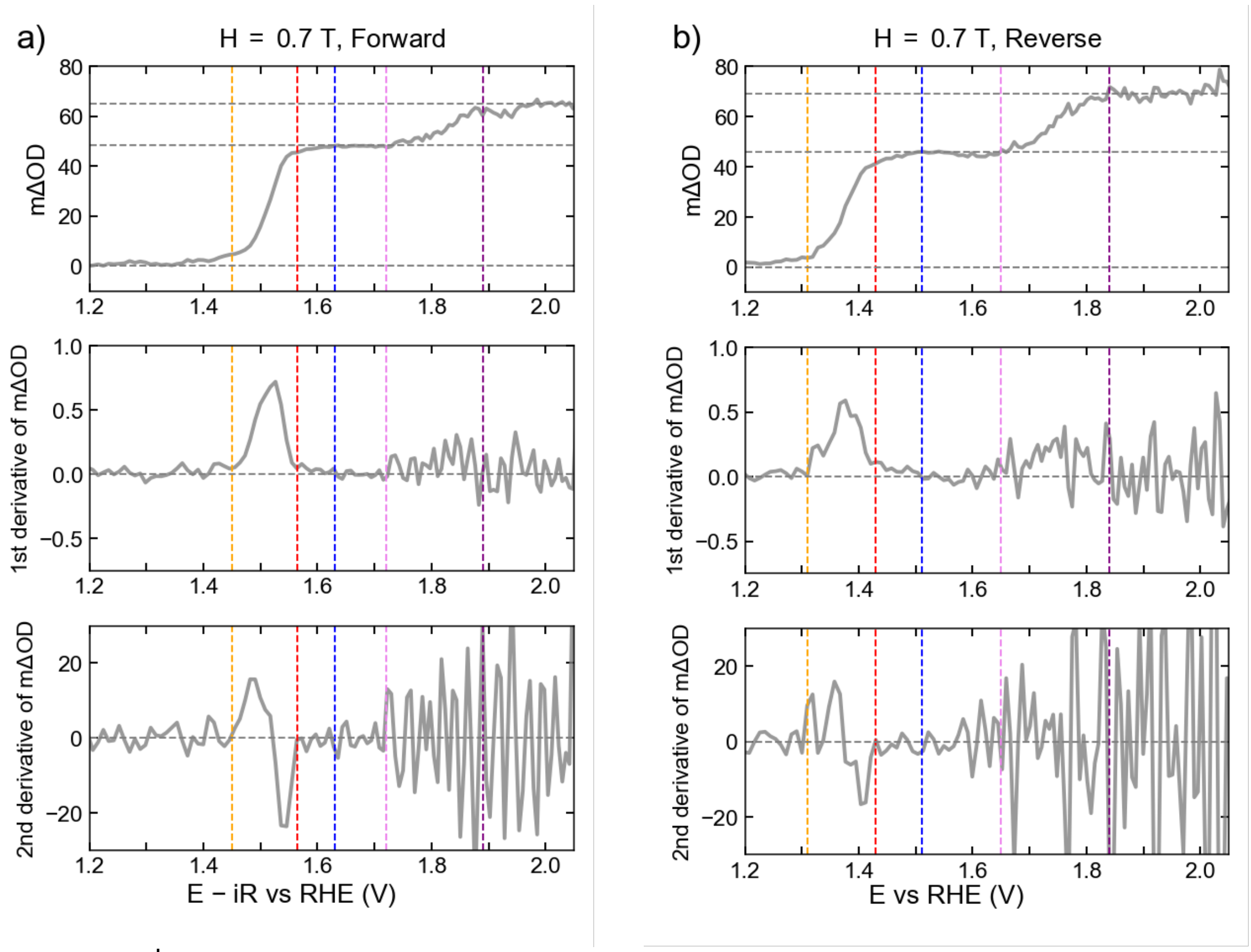

**Figure S14 | Isolation of redox transitions with the derivative method upon magnetic stimulation.** Identifying the 0 and full coverage of redox transitions at H = 0.7 T using the derivative of the optical signal as a function of the potential for the forward or oxidation process (a) and reverse or reduction process (b). Dashed vertical lines indicate the potentials for 0 and full coverage of the redox transitions.

## Supplementary Note 8. Experimental estimates of binding and interaction energies of the redox intermediates

The changes in optical density as a function of applied potential can be used to construct electro-adsorption isotherms. We note that in polaronic systems with strong charge-lattice coupling, such as TMOs, the isotherm model can also be understood as an effective descriptor of the intermediate redox population as opposed to a strict descriptor of the equilibrium state. Yet, this approximation allows assessing parameters, such as the redox potential and interaction energies and has been applied to different metal oxide-based electrocatalysts such as iridium, nickel and cobalt oxides[20,43–46,58]. According to the Lambert-Beer law, the absorbance of an individual redox state at a given potential $\Delta A_E$ is directly proportional to its density D. Thus, the absorbance at saturation $\Delta A_{sat}$, of a redox state is proportional to its full coverage $D_{max}$. Based on this, the coverage θ of each redox state at a given potential is defined as $\theta = D/D_{max} = \Delta A_E/\Delta A_{sat}$. We use the threshold potentials estimated in Supplementary Note 7 as the 0 and full coverage potentials.

The resulting coverage θ vs applied potential, E, data (the electroadsorption isotherm) can be modelled using a simple Langmuir or Frumkin isotherm model:

- Langmuir isotherm:

$$E_{RHE} = E^o_{RHE} + \frac{RT}{nF} ln\left(\frac{\theta}{1-\theta}\right)$$

where RT/nF = 0.0256 V for (n = 1 and T = 295 K). Here, the energy of adsorption is independent of coverage, leading to a Nernst-like dependence and a characteristic thermodynamic potential at $E_{1/2}$ = $E^o_{RHE}$ when $\theta$ = ½. Thus, there are no energetic interactions between the adsorbates.

- Frumkin isotherm:

$$E_{RHE} = E^o_{RHE\ (\theta=0)} + \frac{RT}{nF} ln\left(\frac{\theta}{1-\theta}\right) + \frac{r}{nF}\cdot\theta$$

where $RT/nF$ = 0.0256 V for ($n$ = 1 and T = 295 K) and $r/nF = r$ in the unit of eV. This model is used when the fit with Langmuir isotherm is inadequate, which indicates the non-Nernstian dependence of coverage on potential due to adsorbate interactions[44,46,58,59]. The additional term $\frac{r}{nF}\cdot\theta$ describes the change in adsorption enthalpy as coverage increases, where $r$ is the lateral interaction parameter representing the interaction strength between the adsorbates and $E^o_{RHE\ (\theta=0)}$ is the potential of the redox reaction, assuming no coverage. A positive value of $r$ corresponds to repulsive interactions, while a negative value corresponds to attractive interactions.

The value of $E^o_{RHE}$ or $E^o_{RHE\ (\theta=0)}$ and $r$ can be obtained by fitting the $E$ vs $\theta$ data using the above equations. Adopting a similar approach to that used by Liang et al.[43], we do not include the very initial or final regime of the coverage (< 0.095 or > 0.95) in the fit, because the noise in the optical signal in these regimes makes the detection of the species relatively less reliable. We performed two sets of fits: to the raw data and to the data averaged every 2 to 3 points.

The results of the fittings are shown in **Fig. S15-S18** and **Tables S1-S4** for both the forward and backwards scans. Specifically, **Fig. S15** evaluates the application of the Langmuir isotherm to all the transitions. **Fig. S16** shows the corresponding analysis for data that has been binned every 3 points. Similarly, **Fig. S17** evaluates the application of the Frumkin isotherm to all the transitions. **Fig. S18** shows the corresponding analysis for data that has been binned every 2 points (for 2$^{nd}$ redox transition) or 3 points (for 1$^{st}$ and 3$^{rd}$ redox transitions).

We find that the Langmuir isotherm fits the experimental data well for the Redox Transition 1 at both H = 0 T (red data in **Fig. S15a** and **S15b**) and H = 0.7 T (red data in **Fig. S15c** and **S15d**) and the 3$^{rd}$ redox transition at H = 0.7 T (purple data in **Fig. S15c** and **S15d**). In contrast, the Frumkin isotherm fits the forward scan of Redox Transition 2 at H = 0 T (blue data in **Fig. S15a**) and the forward, and reverse processes of Redox Transition 2 at H = 0.7 T (blue data in **Fig. S17c** and **S17d**) better. We obtain a positive interaction parameter *r* at H = 0 T of ~ 0.29 eV (**Table S3),** similar to observations in other oxides[43,44]. For the case of H = 0.7 T, we observe a small negative *r* parameter of ~ − 0.10 eV (**Table S4**), pointing to an attractive interaction — we note, however, that this transition occurs over a smaller potential range and thus there is a larger uncertainty to its determination and fitting. Thus, further resolution is needed to quantitatively isolate and describe this transition.

We ran a separate analysis with binned data to establish if fluctuations in the signal-to-noise impacted the analysis (**Fig. S16** and **Fig. S18**). We obtain a similar interaction parameter for both the raw and binned data (**Table S1** – **S4**). The good agreement of the experimental data to either the Langmuir or Frumkin models provides further support for the assignments of the redox transitions described in Supplementary Notes 6 and 7. Within the resolution of our experiment and our approach to the isolation of redox transitions, these results suggest that the magnetic field triggers the emergence of a new redox transition and alters adsorbate-adsorbate interactions.

We observed that at H = 0 T, the 2$^{nd}$ redox transition exhibits a more compressed Langmuir-like isotherm during the reverse/reductive CV scan (**Fig. S15b**), while in the forward oxidation scan, the redox transition exhibits a Frumkin isotherm. We attribute the difference in the energetic interactions of oxidised sites for the forward and reverse processes of the 2$^{nd}$ redox transition to the polaronic nature of the active sites with strong charge-lattice coupling. Specifically, the formation of the metal oxo species during the forward redox process entails large structural changes (as evidenced by XAS studies[39,60]), that might lead to stronger adsorbate-adsorbate interactions, as evidenced by the Frumkin behaviour of the coverage revealed by this study.

In contrast, at H = 0.7 T, the forward and reverse processes for each redox transition have similar potential dependence on coverage. This suggests that the applied magnetic field reduces the adsorbate interactions among the oxidised intermediates, especially during the 2$^{nd}$ redox transition or oxo formation step. As suggested by our DFT results, the emergence of the new redox pathways in the magnetically excited NiFeOOH catalysts can alter the energetics of these intermediate states and thereby provide a way to break the linear-scaling relation in this system.

## Langmuir Fits:

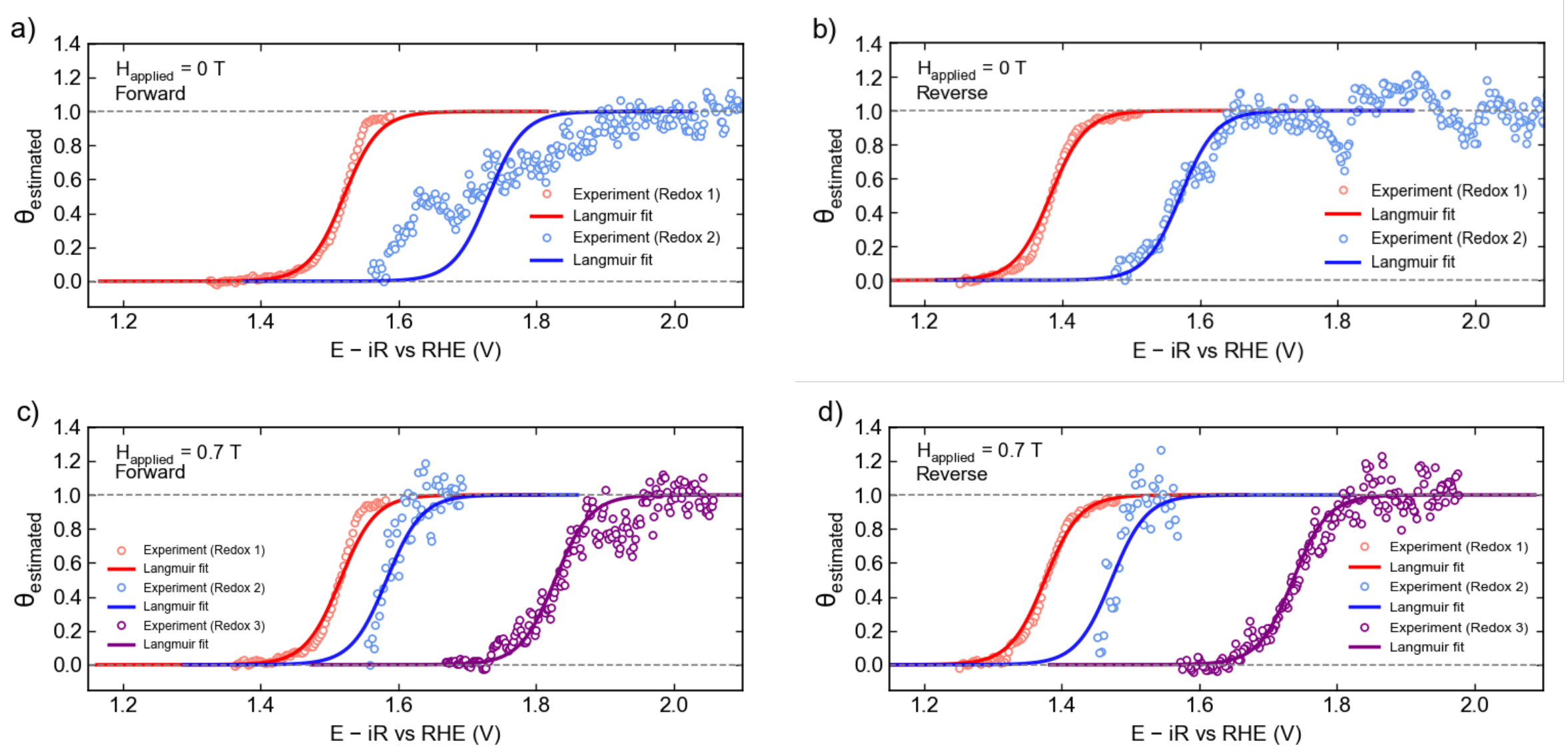

**Figure S15| Langmuir isotherm fitting** at (a,b) H = 0 T and (c,d) H = 0.7 T. In the case of H = 0 T, the Langmuir model could not fit well the experimental coverage data for the 2nd redox transitions at both H = 0 T and H = 0.7 T. A good fit is obtained for all other redox transitions.

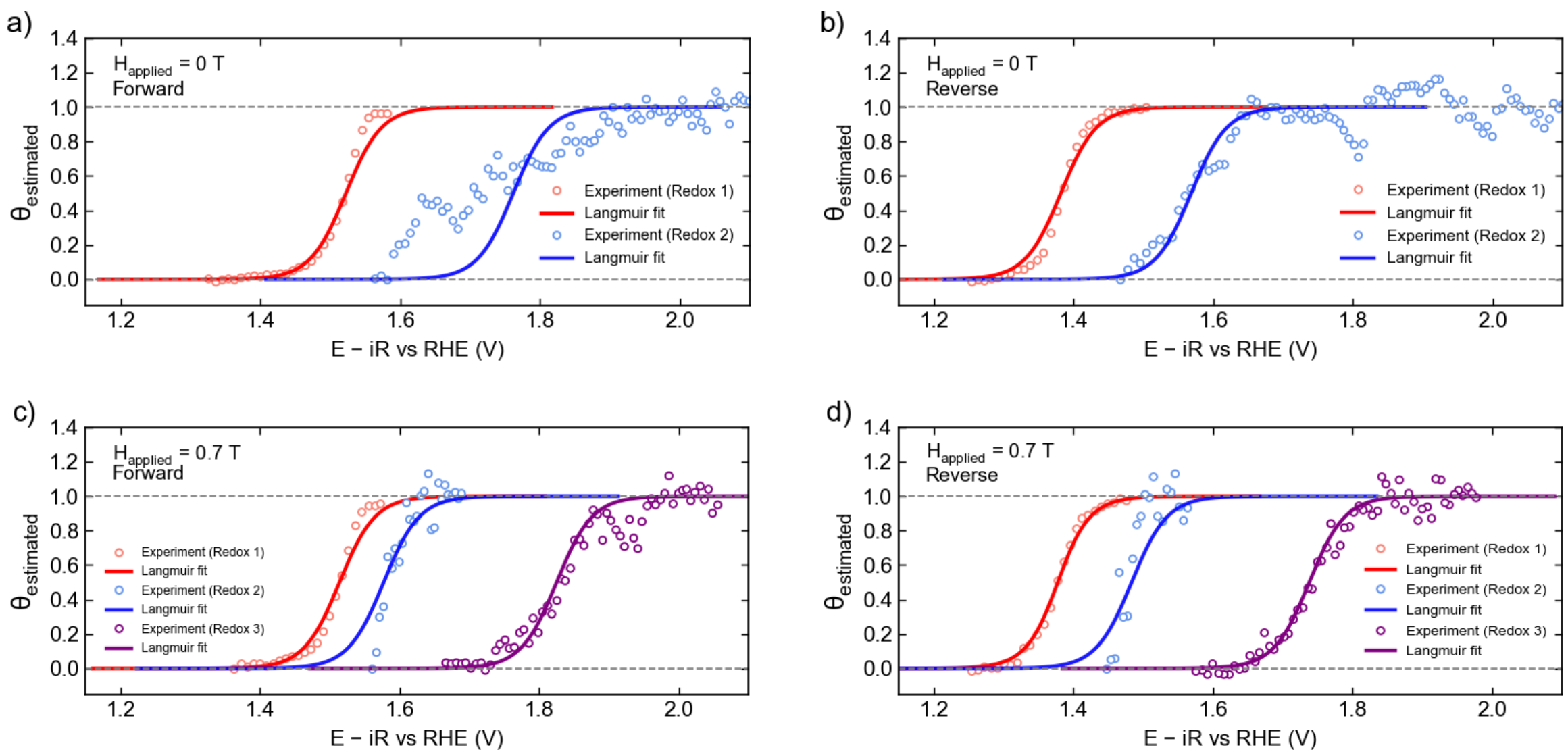

**Figure S16| Langmuir isotherm fitting using binned data** at (a,b) H = 0 T and (c,d) H = 0.7 T. In the case of H = 0 T, the Langmuir model could not fit well the experimental coverage data for the 2nd redox transitions at both H = 0 T and H = 0.7 T. A good fit is obtained for all other redox transitions.

**Table S1|** Summary of parameters obtained from Langmuir Fitting at H = 0 T

| **Quantity** | **Forward** | | **Reverse** | |
|---|---|---|---|---|
| | **Original** | **Binned** | **Original** | **Binned** |
| 1. $E^{o}_{RHE}$ (V) | | | | |
| Redox 1 | 1.52 | 1.52 | 1.38 | 1.38 |
| Redox 2 | 1.73 | 1.76 | 1.57 | 1.57 |

| 2. $R^2$ | | | | |
|---|---|---|---|---|
| Redox 1 | 0.916 | 0.966 | 0.939 | 0.954 |
| Redox 2 | 0.395 | 0.534 | 0.882 | 0.905 |

| Table S2 \| Summary of parameters obtained from Langmuir Fitting at H = 0.7 T | | | | |
|---|---|---|---|---|
| **Quantity** | **Forward** | | **Reverse** | |
| | **Original** | **Binned** | **Original** | **Binned** |
| 1. $E^o_{RHE}$ (V) | | | | |
| Redox 1 | 1.52 | 1.51 | 1.38 | 1.37 |
| Redox 2 | 1.58 | 1.58 | 1.47 | 1.48 |
| Redox 3 | 1.83 | 1.82 | 1.74 | 1.74 |
| 2. $R^2$ | | | | |
| Redox 1 | 0.922 | 0.951 | 0.975 | 0.968 |
| Redox 2 | −1.094 [0.005]* | −0.366 [0.006]* | −2.663 [0.014]* | −3.968 [0.008]* |
| Redox 3 | 0.877 | 0.923 | 0.919 | 0.937 |

*Values in the bracket are $\chi^2$ from the fitting. The very small $\chi^2$ values for Redox 2 indicate that residuals are small and that the Langmuir model can also fit the data well for this transition. The negative $R^2$ can be explained by the systematic deviation of the pattern of residuals from the trend of the data, which in turn, is possible because of the relatively small potential range for this transition in the case of H = 0.7 T. Alternatively, the negative $R^2$ observed for the Langmuir fit of Redox 2 (H=0.7 T) may reflect the inability of a non-interacting model to capture the sharp rise of the magnetically stimulated signal. As shown in Table S4, the fit is significantly improved by using the Frumkin isotherm, suggesting that this transition is driven by attractive adsorbate interactions that violate the Langmuir assumptions.

**Frumkin fits:**

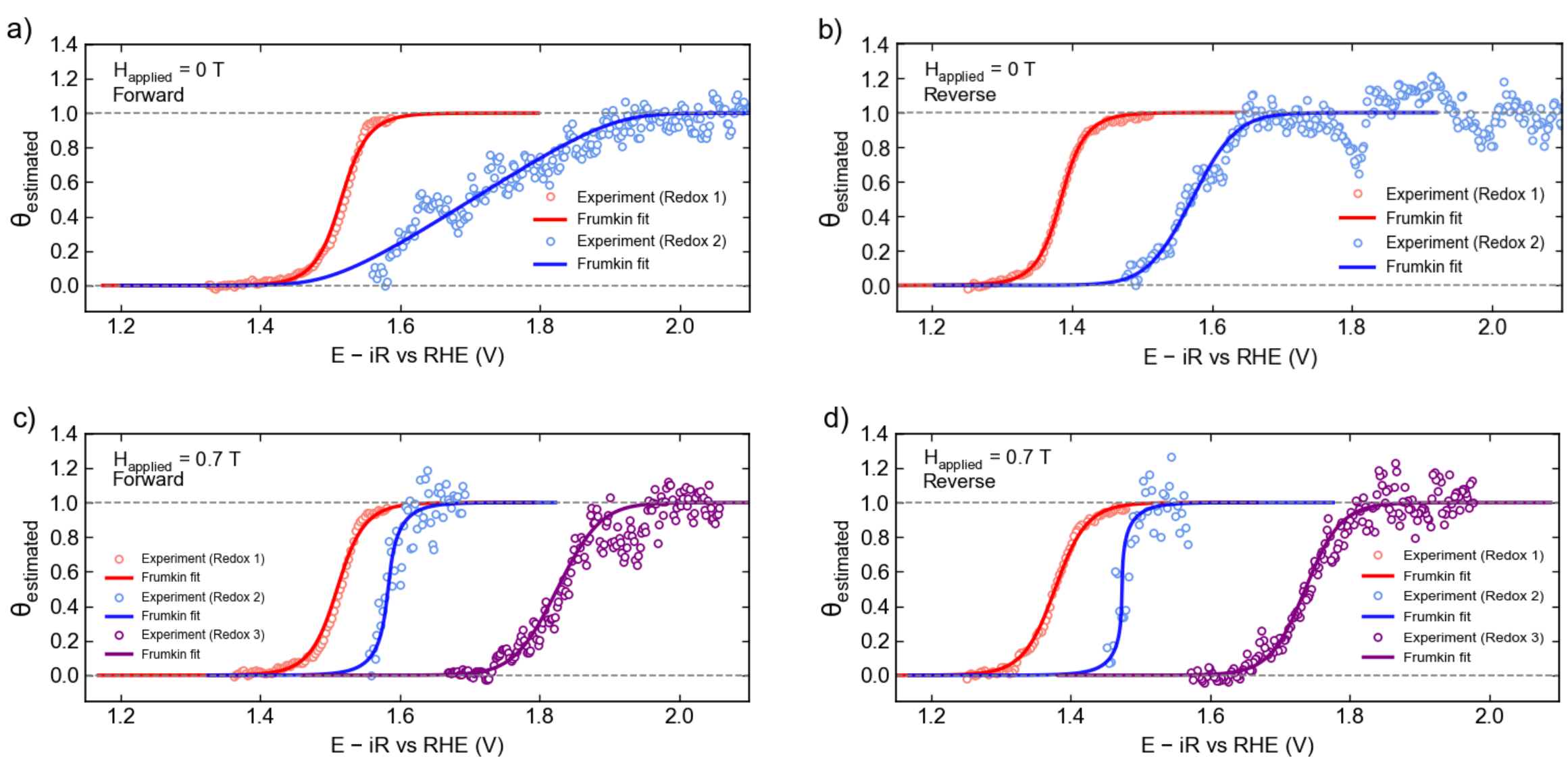

**Figure S17│ Frumkin isotherm fitting** at (a, b) H = 0 T and (c, d) H = 0.7 T. A good fit is obtained for the data of the 2nd redox transition at H = 0 T and H = 0.7 T. At H = 0 T, the fitted interaction energy *r* is positive (r = 0.29 eV), suggesting the adsorbed oxygen species formed at OER potentials exhibit repulsive adsorbate-adsorbate interactions. At H = 0.7 T, the interaction energy becomes small and negative (r = −0.10 eV), suggesting that the magnetic field triggers the suppression of repulsive adsorbate-adsorbate interactions.

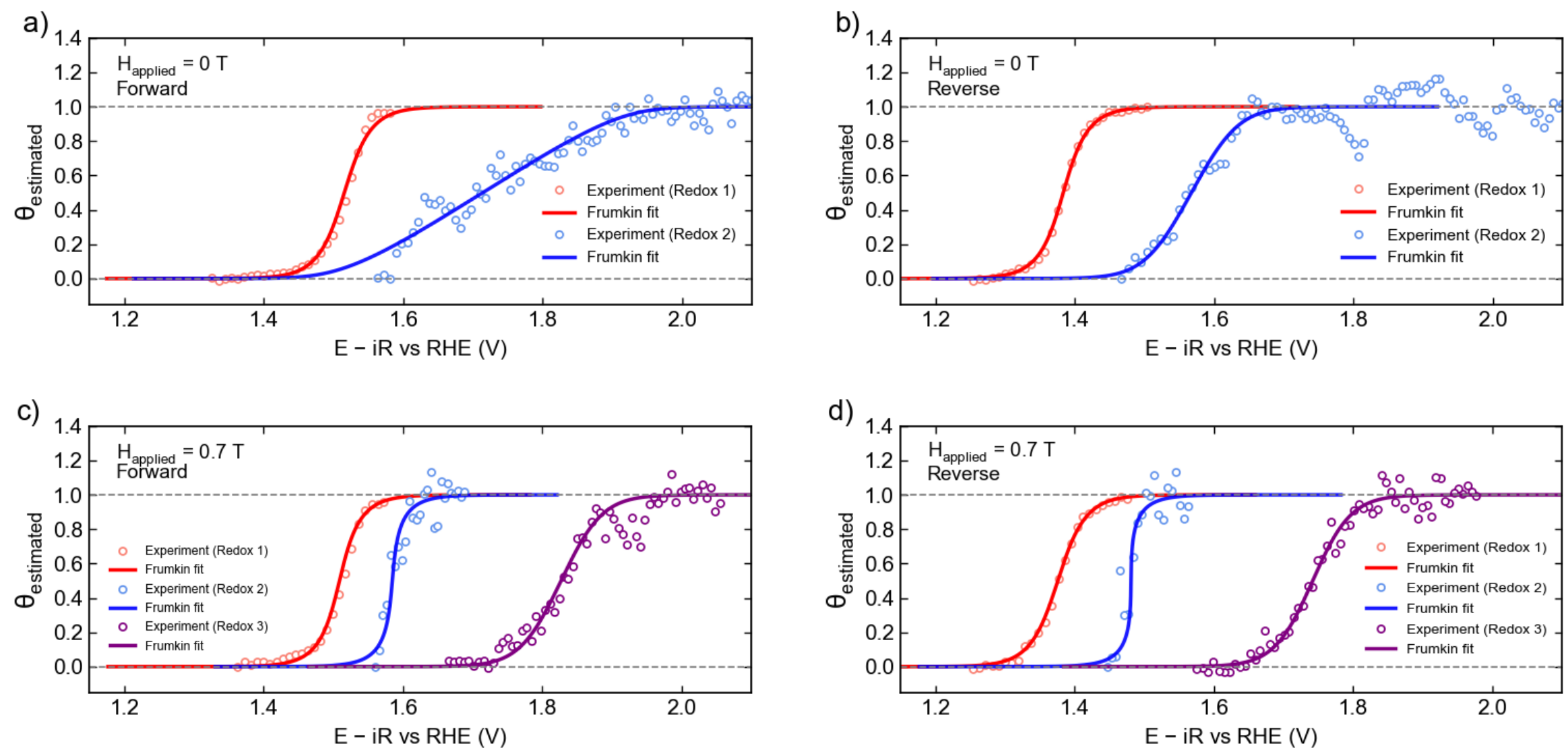

**Figure S18│ Frumkin isotherm fitting using the binned data** at (a, b) H = 0 T and (c, d) H = 0.7 T using the binned experimental data. A good fit is obtained for the data of the 2nd redox transition at H = 0 T and H = 0.7 T. At H = 0 T, the fitted interaction energy *r* is positive (r = 0.29 eV), suggesting the adsorbed oxygen species formed at OER potentials exhibit repulsive adsorbate-adsorbate interactions. At H = 0.7 T, the interaction energy becomes small and negative (r = − 0.10 eV), suggesting that the magnetic field triggers the suppression of repulsive adsorbate-adsorbate interactions although the sharpness of this transition means there are significantly less points increasing the uncertainty in its isolation and quantification (see discussion in the text).

**Table S3│** Summary of parameters obtained from Frumkin Fitting at H = 0 T

| **Quantity** | **Forward** | | **Reverse** | |
|---|---|---|---|---|
| | **Original** | **Binned** | **Original** | **Binned** |
| 1. $E^{o}_{RHE\ (\theta=0)}$ (V) | | | | |
| Redox 1 | 1.53 | 1.53 | 1.40 | 1.40 |
| Redox 2 | 1.56 | 1.57 | 1.56 | 1.55 |
| 2. *r* (eV) | | | | |
| Redox 1 | −0.02 | −0.03 | −0.03 | −0.03 |
| Redox 2 | 0.30 | 0.29 | 0.03 | 0.03 |
| 3. $R^2$ | | | | |
| Redox 1 | 0.979 | 0.974 | 0.994 | 0.994 |
| Redox 2 | 0.832 | 0.878 | 0.906 | 0.945 |

**Table S4│** Summary of parameters obtained from Frumkin Fitting at H = 0.7 T

| **Quantity** | **Forward** | | **Reverse** | |
|---|---|---|---|---|
| | **Original** | **Binned** | **Original** | **Binned** |
| 1. $E^{o}_{RHE\ (\theta=0)}$ (V) | | | | |
| Redox 1 | 1.53 | 1.53 | 1.38 | 1.39 |
| Redox 2 | 1.62 | 1.63 | 1.52 | 1.53 |
| Redox 3 | 1.81 | 1.82 | 1.73 | 1.74 |
| 2. *r* (eV) | | | | |
| Redox 1 | −0.04 | −0.04 | −0.01 | −0.02 |

| Redox 2 | −0.07 | −0.08 | −0.10 | −0.10 |
| --- | --- | --- | --- | --- |
| Redox 3 | 0.02 | 0.01 | 0.02 | 0.01 |
| 3. $R^2$ | | | | |
| Redox 1 | 0.982 | 0.985 | 0.986 | 0.994 |
| Redox 2 | 0.661 | 0.707 | 0.692 | 0.938 |
| Redox 3 | 0.858 | 0.929 | 0.923 | 0.959 |